\documentclass[a4paper,USenglish,numberwithinsect,cleveref,autoref]{lipics-v2021}
\usepackage{mathtools}
\usepackage{booktabs}
\usepackage{xspace}
\usepackage{float}
\usepackage{placeins}
\usepackage{needspace}
\usepackage{algorithm}
\usepackage[noend]{algpseudocode}
\crefname{algorithm}{Algorithm}{Algorithms}
\Crefname{algorithm}{Algorithm}{Algorithms}
\makeatletter
\renewcommand*{\theHALG@line}{\thealgorithm.\arabic{ALG@line}}
\makeatother

\newcommand{\eps}{\varepsilon}
\newcommand{\wtO}{\widetilde O}
\newcommand{\one}{\mathbf 1}
\newcommand{\TSP}{\operatorname{TSP}}
\newcommand{\HK}{\operatorname{HK}}
\newcommand{\cng}{\mathsf{cong}}
\newcommand{\mincut}{\operatorname{mincut}}
\newcommand{\Focus}{\textsf{Focus}\xspace}
\newcommand{\StaticExtract}{\textsf{FindCuts}\xspace}
\newcommand{\StreamSparsify}{\textsf{StreamSparsify}\xspace}
\newcommand{\poly}{\operatorname{poly}}
\newcommand{\Good}{\mathcal E_{\mathrm{good}}}
\newcommand{\CapsPoly}{\mathsf P_{\log}}
\newcommand{\Est}{\widehat{\mathsf T}}
\newcommand{\nph}{N_{\rm ph}}

\title{Semi-Streaming Estimation of the Held--Karp Bound and Graph-Metric TSP Cost}
\titlerunning{Semi-Streaming Held--Karp and TSP Cost Estimation}
\hideLIPIcs
\author{Sharareh Alipour}{Tehran Institute for Advanced Studies}{}{}{}
\authorrunning{Sharareh Alipour}
\Copyright{Sharareh Alipour}
\AtBeginDocument{\hypersetup{pdfauthor={Sharareh Alipour},pdftitle={Semi-Streaming Estimation of the Held--Karp Bound and Graph-Metric TSP Cost},pdfsubject={Full manuscript}}}

\nolinenumbers

\begin{document}
\maketitle

\begin{abstract}
The Held--Karp bound is a linear-programming lower bound for the metric traveling salesperson problem (TSP). We approximate this bound in the multi-pass insertion-only streaming model for metrics induced by connected positive-weight graphs. For an $n$-vertex, $m$-edge graph and $0<\eps\le1/2$, the algorithm returns a $(1+\eps)$ upper approximation with high probability, using $\wtO(\eps^{-4})$ passes, $\wtO(n\eps^{-2})$ words, and $\wtO(m\eps^{-4})$ arithmetic operations per pass.

We replace the changing edge-weight vector by a short history of multiplicative-weights (MWU) updates. After initialization, updates repeat until a column is removed. We store each repeated update once, with its repetition count. The record length depends on the number of columns, not updates; the argument applies to any fixed nonnegative matrix. For Held--Karp, tree-path queries recover edge weights from these records. Separate search and update thresholds allow fresh sparsifiers at successive search stages. Controlled rounding gives a bounded-precision implementation.

The estimate yields a $(3/2+\eps)$ graph-metric TSP value estimate and a $(7/5+\eps)$ estimate when the input graph is unweighted. Conversely, in weighted graph streams, every one-pass $(1+\eta)$ upper approximation to Held--Karp, for fixed $0\le\eta<1$, requires quadratic space. The algorithms estimate values; they do not construct tours within these bounds.
\end{abstract}

\clearpage

\section{Introduction}
Let $G=(V,E)$ be a connected undirected graph with positive edge
costs $c=(c_e)_{e\in E}$, $n:=|V|$, and $m:=|E|$. Let $\TSP(G)$
be the minimum tour cost in its shortest-path metric, equivalently
the minimum cost of a closed walk visiting every vertex of $G$.
For $S\subseteq V$, let $\delta_G(S)$ contain the edges with exactly one endpoint in $S$.
The Held--Karp bound is the subtour-elimination LP optimum in the
complete shortest-path metric. We use its equivalent cut-covering
formulation on the input graph, where $y_e$ is the fractional
multiplicity of edge $e$~\cite{GoemansBertsimas,MonmaMunsonPulleyblank}:

\[
\begin{aligned}
 \HK(G)=\min\quad &\sum_{e\in E}c_ey_e\\
 \text{subject to}\quad &y_e\ge0 &&(e\in E),\\
 &\sum_{e\in\delta_G(S)}y_e\ge2
       &&(\varnothing\subsetneq S\subsetneq V).
\end{aligned}
\]
Unlike the complete-metric formulation, this LP has no degree-two
equations. We call a feasible vector a \emph{fractional cover}; it need
not describe a tour. Every closed spanning walk crosses each nontrivial
cut at least twice, so $\HK(G)\le\TSP(G)$.

We work in the \emph{multi-pass insertion-only semi-streaming model}. The input is a read-only list of weighted edges, with no deletions. A \emph{pass} is one complete scan of this list; the arbitrary input order and edge costs remain fixed across passes. The algorithm uses $\wtO(n\poly(1/\eps))$ words of memory, where $\eps$ is the requested accuracy.

Chekuri and Quanrud~\cite{ChekuriQuanrud} give a $(1+\eps)$
approximation in $O(m\log^4 n/\eps^2)$ time, with high probability.
Their method repeatedly finds small cuts under auxiliary edge weights
and updates those weights; the input costs stay fixed. Delayed updates
save time, but storing information for every edge exceeds
semi-streaming space on dense graphs.

Passes pose a second difficulty. Each cut search depends on earlier
updates, and the sequential process can require $\Omega(m)$
updates~\cite{KohWeinsteinYing}. One scan per update would therefore
use too many passes. Koh et al.~\cite{KohWeinsteinYing} process groups
of cuts instead, obtaining $(1+\eps)$ accuracy with nearly linear work
and polylogarithmic parallel depth. Their algorithm still stores a
weight for every edge.

\subsection{Our results}
We store a short update history and reconstruct edge weights during
each pass. Cut searches use a fresh \emph{sparsifier}: a small reweighted
graph that approximately preserves the current cut values. It is a
search tool, not a replacement LP (\cref{rem:no-static-coreset}).\footnote{$\wtO$ hides polylogarithms in $n,m,1/\eps$ and the cost ratio;
Section~\ref{sec:prelim} defines word sizes.}

\begin{theorem}[Semi-streaming Held--Karp estimation]\label{thm:main-HK}
For every connected graph $G$ with positive polynomial-bit edge
costs and $0<\eps\le1/2$, there is a randomized algorithm in the
multi-pass insertion-only streaming model. On every run, it uses
$\wtO(\eps^{-4})$ passes, $\wtO(n\eps^{-2})$ words, and
$\wtO(m\eps^{-4})$ arithmetic operations per pass.

Let $\Good$ be the event that every randomized procedure satisfies
its guarantee. This event has high probability. On $\Good$, the
algorithm returns $\widehat H$ satisfying
$\HK(G)\le\widehat H\le(1+\eps)\HK(G)$.
Outside $\Good$, it may return \textnormal{\textsc{Fail}} or an
incorrect number.
\end{theorem}

On the success event, a saved history prefix recovers a feasible fractional cover whose cost is at most the output.
Our Held--Karp estimate gives the following TSP cost guarantees.

\begin{corollary}[TSP cost estimation]\label{cor:tsp}
Under the assumptions of \cref{thm:main-HK}, a randomized algorithm
returns, with high probability, an estimate $\Est$ satisfying
$\TSP(G)\le\Est\le(3/2+\eps)\TSP(G)$.
For unweighted graphs, the approximation factor improves to $7/5+\eps$.
Both guarantees use the pass, space, and work bounds of
\cref{thm:main-HK}.
\end{corollary}

For graph metrics, add half the Held--Karp estimate to the cost of a
minimum spanning tree (MST). For unweighted graphs, multiply the estimate
by $7/5$, using Seb\H{o}--Vygen's Theorem~29~\cite{SeboVygen}.
More generally, a bound $\TSP\le\gamma\HK$ gives a
$\gamma(1+\eps)$ estimate. Adjusting the internal accuracy gives
\cref{cor:tsp}.

Multiple passes are necessary for our accuracy guarantee. We show that,
for every fixed $0\le\eta<1$, a one-pass $(1+\eta)$-upper approximation
to the Held--Karp value requires $\Omega(n^2)$ bits on weighted graph
streams.

\subsection{Main ideas}
Three ideas let the algorithm maintain changing edge weights without
storing either the full vector or every update.

\emph{Record repeated updates once.} Within a cut group, each update
after initialization is identical until a cut reaches its target and
leaves. Store that update with its repetition count. Since a cut leaves
at most once, record length depends on the number of cuts, not updates.
This argument uses only the update rule, not graph structure.

\emph{Recover weights as edges arrive.} A cut is described by one or two
edges of a stored tree~\cite{MukhopadhyayNanongkai}. Tree-path queries
identify the recorded updates affecting an arriving edge. Its recovered
weight is used and then discarded. The same queries compute cut values
and the next update size, so the record supports the continuing algorithm.

\emph{Allow for search error.} A fresh sparsifier may underestimate a
previously processed cut. We raise selected cuts above the search
threshold and lower that threshold slightly between stages. This gap
absorbs the error and preserves earlier progress.

\subsection{Related work}
\Cref{tab:comparison} compares the closest streaming results; all allow
arbitrary edge order. To the best of our knowledge, ours is the first
semi-streaming algorithm that directly approximates Held--Karp.

\begin{table}[!htb]
\caption{Streaming results; space in words. Outputs are randomized
cost estimates, except for MST doubling (value or tour), the tours
in~\cite{AlipourFarokhnejadMomke}, and the deterministic $1.96$
estimate~\cite[Theorem~6]{ChenKhannaTan}. For the one-pass $1.83$
implementation, see~\cite[p.~37:3, footnote~1]{ChenKhannaTan}.}
\label{tab:comparison}
\centering
\begin{tabularx}{\textwidth}{@{}l>{\raggedright\arraybackslash}Xccl@{}}
\toprule
Problem & Source & Factor & Passes & Space\\
\midrule
Graph-metric TSP & MST doubling~\cite{Schrijver}
 & $2$ & $1$ & $\wtO(n)$\\
Graphic TSP & Chen--Kannan--Khanna~\cite{ChenKannanKhanna}
 & $11/6$ & $1$ & $\wtO(n)$\\
Graphic TSP & Behnezhad et al.~\cite{BehnezhadEtAl}
 & $1.83$ & $1$ & $\wtO(n)$\\
Graph-metric TSP & Chen--Khanna--Tan~\cite{ChenKhannaTan}
 & $1.96$ & $2$ & $\wtO(n)$\\
$(1,2)$-TSP & Alipour et al.~\cite{AlipourFarokhnejadMomke}
 & $4/3+\eps$ & $\poly(1/\eps)$ & semi-streaming\\
\midrule
Held--Karp & This work
 & $1+\eps$ & $\wtO(\eps^{-4})$ & $\wtO(n\eps^{-2})$\\
Graph-metric TSP & This work
 & $3/2+\eps$ & $\wtO(\eps^{-4})$ & $\wtO(n\eps^{-2})$\\
Graphic TSP & This work
 & $7/5+\eps$ & $\wtO(\eps^{-4})$ & $\wtO(n\eps^{-2})$\\
\bottomrule
\end{tabularx}
\end{table}

Mukhopadhyay--Nanongkai compute an exact weighted minimum cut in $O(\log n)$ passes and $\wtO(n)$ space using cuts that cross at most two edges of a reference tree~\cite{MukhopadhyayNanongkai}. Other work studies one-pass minimum cut, connectivity augmentation, and network design~\cite{DingEtAl,JinEtAl,ChekuriJainMahabadiVakilian}. 

Our algorithm uses multiple passes to find small cuts under changing
auxiliary capacities without storing the full capacity vector.
It does not directly yield a sublinear-query algorithm for metrics
given by distance queries.

\emph{Organization.}
Sections~\ref{sec:prelim}--\ref{sec:transfer} give the LP and TSP
reductions. Section~\ref{sec:static} gives the cut-search and update
procedures, Section~\ref{sec:replay} compresses their history, and
Section~\ref{sec:buffer} proves progress across searches.
Section~\ref{sec:assembly} assembles and analyzes the algorithm.
The appendices supply source details, rounding, examples, and the
TSP transfer proofs.

\section{Preliminaries}\label{sec:prelim}

We discard self-loops and write $\delta(S)$ for $\delta_G(S)$
when the graph is clear. A cut is nontrivial if
$\varnothing\subsetneq S\subsetneq V$; $S$ is its shore.
For an edge vector $v$ and an edge set $F$, write
$v(F):=\sum_{e\in F}v_e$.

We use $\log$ for the natural logarithm, $\langle u,v\rangle$ for the
inner product, and $\|v\|_\infty:=\max_i|v_i|$.
The vector $x|_J$ agrees with $x$ on the coordinates in $J$ and is
zero elsewhere.
For $n=1$, $\TSP(G)=\HK(G)=0$; for $n=2$, both equal twice
the minimum edge cost.

\subsection{The normalized LP}
Let $z:=\HK(G)/2$. Scaling the input-graph formulation by one half gives
\begin{equation}\label{eq:unit-cover}
 z=\min\left\{\sum_{e\in E}c_ey_e:y\ge0,\ y(\delta(S))\ge1
                    \text{ for every nontrivial cut }S\right\}.
\end{equation}
To count each cut once, fix a root $r_0$ and let $\mathcal C$ contain all nonempty $S\subseteq V\setminus\{r_0\}$. The dual gives each cut a weight $x_S\ge0$, subject to $\sum_{S:e\in\delta(S)}x_S\le c_e$ for every edge.

Let $\one$ be an all-ones vector. Define $A_{e,S}=1/c_e$ when
$e\in\delta(S)$, and zero otherwise. Dividing each dual constraint
by $c_e$ and substituting $w_e=c_ey_e$ in the cover gives the
normalized packing and covering LPs:
\[
 \max\{\one^\top x:Ax\le\one,\ x\ge0\},\qquad
 \min\{\one^\top w:A^\top w\ge\one,\ w\ge0\}.
\]
Both optima equal $z$ by strong duality. The algorithm maintains
positive auxiliary weights $w$, not necessarily a feasible cover.
The input costs $c_e$ stay fixed. The current \emph{capacities} are
$q_e:=w_e/c_e$, the total weight is $D:=\sum_e w_e$, and a cut's
current value is $(A^\top w)_S=q(\delta(S))$.

Write $\mincut(q):=\min_{S\in\mathcal C}q(\delta(S))$.
If $0<\lambda\le\mincut(q)$, then $y_e=q_e/\lambda$
is a feasible demand-one cover of cost $D/\lambda$.
The algorithm maintains such a lower bound $\lambda$ and saves
the smallest ratio $D/\lambda$ seen.

For a stored graph $H$, let $q_H(F)$ denote its total stored
weight on $F\subseteq E(H)$, and let $\mincut(H)$ denote its
minimum-cut value.

\subsection{Arithmetic}

Let $\ell_c$ be the maximum input-cost encoding length.
A word contains
$\Theta(\log n+\log m+\ell_c+\log(1/\eps))$ bits.
Our space bounds count words; multiplying by the word length gives
the bit-space bound. For polylogarithmic-bit costs, this is the
usual semi-streaming convention.
``With high probability'' means with probability at least
$1-n^{-c_0}$ for any fixed $c_0>0$.

We first use exact arithmetic, with internal accuracy $\xi$ chosen
proportional to $\eps$ in Section~\ref{sec:assembly}.
Appendix~\ref{app:precision} gives a bounded-precision implementation
with the same approximation and asymptotic resource bounds, and
explains how to handle zero-cost edges.

\section{TSP bounds and the one-pass lower bound}\label{sec:transfer}

Let $W(G)$ be the minimum-spanning-tree cost, abbreviated to $W$. The following bound gives both a TSP estimate and a one-pass space lower bound.

\begin{theorem}[MST and Held--Karp bounds]\label{thm:MST-sandwich}
For every connected positive-cost graph, $\TSP(G)\le W(G)+z\le\tfrac32\TSP(G)$.
\end{theorem}

The proof is given in Appendix~\ref{app:tsp-transfer}.

\begin{corollary}[MST transfer]\label{cor:MST-transfer}
If $\eta\ge0$ and $z\le Z\le(1+\eta)z$, then $\TSP(G)\le W(G)+Z\le(\tfrac32+\tfrac\eta2)\TSP(G)$.
Also, $2Z$ is a $(1+\eta)$ upper approximation to $\HK(G)$.
\end{corollary}

\begin{proof}
The lower bound follows from \cref{thm:MST-sandwich}.
For the upper bound, $W(G)+Z\le\TSP(G)+(1+\eta)\TSP(G)/2$.
The final claim follows from $\HK(G)=2z$.
\end{proof}

\begin{proposition}[One pass is insufficient in weighted graph streams]
\label{prop:one-pass-lb}
Let $0\le\eta<1$. Suppose a randomized one-pass algorithm uses
$s_{\rm HK}$ bits and, on every connected positive-cost graph
stream, returns $\HK(G)\le\widehat H\le(1+\eta)\HK(G)$
with probability at least $2/3$. Then $s_{\rm HK}=\Omega((1-\eta)^2n^2)-O(n\log n)$.
For each fixed $\eta<1$, this gives $\Omega(n^2)$ bits.
The claim concerns weighted streams, not unweighted input graphs.
\end{proposition}

\begin{proof}
Chen--Khanna--Tan's Theorem~5~\cite{ChenKhannaTan} requires $\Omega(\vartheta^2n^2)$ bits for a one-pass $(2-\vartheta)$ approximation to TSP cost. Theorem~13~\cite[Part~I, Section~4.2]{ChenKhannaTanFull} gives the hard instances: connected graphs with polynomially bounded positive integer costs.
An exact MST can therefore be maintained in one pass using
$O(n\log n)$ bits on these instances.

Run this MST computation alongside the Held--Karp estimator and
output $\Est_0:=W(G)+\widehat H/2$.
By \cref{cor:MST-transfer}, on success, $\TSP(G)\le\Est_0\le\alpha\TSP(G)$, where $\alpha:=\tfrac32+\tfrac\eta2 =2-\tfrac{1-\eta}{2}$. Their theorem is stated for estimates below the optimum. Set $Y:=\Est_0/\alpha$ to obtain $Y\le\TSP(G)\le\alpha Y$.
A constant number of independent copies and a median reduce the
failure probability below $1/10$ without adding passes.
Applying the lower bound with $\vartheta=(1-\eta)/2$ and
subtracting the MST storage proves the claim.
\end{proof}

\begin{theorem}[Using a stronger tour bound]\label{thm:LP-transfer}
Suppose $\TSP(G)\le\gamma\HK(G)$ for some $\gamma\ge1$. For $\eta\ge0$ and $z\le Z\le(1+\eta)z$,
the estimate $\Est_\gamma:=\min\{W(G)+Z,2\gamma Z\}$
satisfies $\TSP(G)\le\Est_\gamma\le\min\{\tfrac32+\tfrac\eta2,\gamma(1+\eta)\}\TSP(G)$.
\end{theorem}

The MST term can improve the estimate: $W+z$ is exact on trees,
although its ratio approaches $3/2$ on unit-cost cycles.
Appendix~\ref{app:tsp-transfer} gives these examples, the proof of
\cref{thm:LP-transfer}, and the rounding needed for \cref{cor:tsp}.

\section{Cut searches and group updates}\label{sec:static}

We adapt two procedures from Koh et al. (KWY)~\cite{KohWeinsteinYing}:
$\StaticExtract$ finds cuts below a threshold in a stored graph;
$\Focus$ updates the selected cut variables. We state their guarantees
here and implement the updates in the stream in Section~\ref{sec:replay}.
Appendix~\ref{app:static-contract} checks the search assumptions;
Table~\ref{tab:kwy-import} lists the source results. All KWY locators
refer to the cited arXiv version.

\subsection{Tree cuts and search order}
Fix a spanning tree $T$ rooted at $r_0$. We call it a
\emph{reference tree} because we use its edges to describe cuts.
For a cut $S$, write $\delta_T(S):=\delta_G(S)\cap E(T)$.
When this set contains one or two edges, we store those edges
instead of the vertices of $S$.
For a tree edge $a$, let $V_a$ be its descendant vertex set.
Then $\{a\}$ describes $V_a$ and $\{a,b\}$ describes
$V_a\triangle V_b$, where $\triangle$ denotes symmetric difference.

\begin{theorem}[{\cite[Theorem~2.3]{KohWeinsteinYing}}]\label{thm:tree-packing}
For a connected positive-weight graph $H$, one can compute
$O(\log n)$ spanning trees using $\wtO(|E(H)|+n)$ work and storage
such that, with high probability, every cut of weight at most
$\tfrac54\mincut(H)$ crosses at most two edges of at least one tree.\end{theorem}

Searches follow a fixed order: one tree edge, two edges on the same
decomposition path, then two edges on different paths. The tree fixes
a cut's stage; current capacities determine whether it is found.

\begin{definition}[Cut-search stages]\label{def:stage-families}
At each nonleaf vertex, mark the edge to a child with a largest
subtree as \emph{heavy}, breaking ties by a fixed rule.
Follow heavy edges from the root to form one path.
Each nonheavy edge starts another path, continued along heavy
edges. The nonempty paths partition $E(T)$, and every root-to-leaf
path meets $O(\log n)$ of them.

Recursively split each path's edge sequence into consecutive halves,
putting any extra edge in the first half, until single edges remain.
These intervals form a binary tree: single-edge leaves have height
zero, and each parent has height one plus the larger child height.

A cut crossing one tree edge belongs to stage zero.
A cut crossing two edges on the same path belongs to the stage
given by the height of the smallest recursive interval containing
both edges. All pairs on different paths belong to one final
stage, one above the largest interval height.
Let $\mathcal F_j(T)$ be the cuts assigned to stage $j$.
There are at most $\lceil\log_2(n-1)\rceil+2$ stages.
These families depend only on $T$ and are not listed in full.
\end{definition}

A \emph{stage} searches one cut family and may update the cuts it
finds. An \emph{outer round} processes all stages of several trees to
increase the minimum-cut lower bound (Section~\ref{sec:buffer}).
Neither is a single pass. Appendix~\ref{app:stage-example} illustrates
the stages; Appendix~\ref{app:example} follows a complete outer round.

\begin{lemma}[Earlier stages supply the path-search assumptions]\label{lem:root-prerequisites}
Let $Q$ be an interval of height $j$ with child intervals $Q_L,Q_R$.
After contracting all tree edges outside $Q$, write the remaining
path as $v_0,e_1,v_1,\ldots,e_k,v_k$. If $Q_L$ ends with $e_h$
and $Q_R$ starts with $e_{h+1}$, call $v_h$ their separator.
For $p<q$, the middle shore of $\{e_p,e_q\}$ is
$\{v_p,\ldots,v_{q-1}\}$. It contains $v_h$ exactly when
$e_p$ and $e_q$ lie in different children of $Q$, precisely the
pairs assigned to stage $j$. Otherwise both edges lie in one child
and were assigned earlier. Expanding each contracted vertex back
to its original vertices gives the same cut value in the original graph.
\end{lemma}

\begin{proof}
The middle shore contains $v_h$ exactly when $p\le h<q$.
Otherwise both edges lie in one child, whose height is smaller than
$j$. Contraction preserves the corresponding cut and its value.
\end{proof}

KWY~\cite[Lemma~4.9 and Definition~4.16]{KohWeinsteinYing}
call a pair \emph{root-containing} when its middle shore contains
the separator, and \emph{nonroot} otherwise; this is unrelated to
the global root $r_0$.

\begin{proposition}[Cut-search guarantee]\label{prop:static-package}
\textnormal{(Adapted from~\cite[Lemmas~4.8, 4.9, 4.24]{KohWeinsteinYing}.)}
Fix a positive-weight graph $H$ containing $T$, a stage $j$,
and $\theta>0$. If every cut assigned to an earlier stage has
weight at least $\theta$ in $H$, then
$\StaticExtract(H,T,\theta,j)$ returns exactly
$B_j=\{S\in\mathcal F_j(T):q_H(\delta(S))<\theta\}$.
This group contains $\wtO(n)$ cuts.
The procedure uses $\wtO(|E(H)|+n)$ work and storage, with
limits enforced on every run. The stage families depend only
on $T$, while $B_j$ depends on $H$. No relation between
$\theta$ and $\mincut(H)$ is required.
\end{proposition}

Stage zero checks one-edge cuts directly. For same-path pairs,
\cref{lem:root-prerequisites} places every nonroot pair required by
KWY's Lemma~4.9~\cite{KohWeinsteinYing} in an earlier stage.
The final stage uses KWY's Lemma~4.24 to select path pairs, after all
one-edge and same-path cuts meet the threshold.

The source procedures may also list cuts assigned elsewhere. We keep
only stage-$j$ cuts and check their weights in $H$ using
\cref{prop:range-oracle}(ii). Appendix~\ref{app:static-contract}
proves that these searches miss no required cut, return $\wtO(n)$
candidates, and allow any $\theta>0$.

\subsection{Updating one cut group}
Given a cut group $B$, $\Focus(B,U)$ updates weights until every
selected cut reaches $U$ or the global load limit is reached.
For the general matrix rule below, columns are variables and rows are
constraints; in our LP they are cuts and edges. An update $g$ adds
load $(Ag)_e$ to row $e$. Its \emph{width}, $\|Ag\|_\infty$, is the
largest added load.

\begin{proposition}[Group-update rule and iteration bound]\label{prop:focus-interface}
\textnormal{(Adapted from Algorithm~2 and Lemma~3.6~\cite{KohWeinsteinYing}.)}
Let $A$ be fixed and nonnegative, with $m\ge2$ rows. For
$0<\xi<1/10$, set $\Gamma:=(\log m)/\xi$. Let $w>0$ be the
current row weights and $\cng$ the sum of previous row loads.
For a target $U>0$, a column $S$ is \emph{active} while
$(A^\top w)_S<U$.

An update $g$ changes $w_e\leftarrow w_e(1+(Ag)_e)$ and
$\cng\leftarrow\cng+Ag$. After each update, remove from the active
group all columns whose values reach $U$.
For a group $B$ of $b\ge1$ nonzero active columns, set
$\alpha_S:=\xi/(b\max_e A_{e,S})$ on $B$ and zero elsewhere.
Initialize $x:=\alpha$ and apply $g=\alpha$ as the first update.
While the remaining active group $\mathcal A$ is nonempty and
$\|\cng\|_\infty<\Gamma$, apply
$g:=\xi x|_{\mathcal A}/\|A(x|_{\mathcal A})\|_\infty$ and set
$x\leftarrow x+g$. Thus the local vector $x$ records only this group's updates.

Return before initialization if $B$ is empty or the load limit is
already reached. Otherwise the call takes at most
\[
 J_b:=2+\left\lceil\frac{8\Gamma}{\xi}
       \log\!\left(2+\frac{2b\Gamma}{\xi}\right)\right\rceil
\]
updates, including initialization. The logarithmic dependence on $b$ is retained for a general matrix.
For the cut groups of \cref{prop:static-package}, $b=\wtO(n)$ and
$J_b=\wtO(\xi^{-2})$.
\end{proposition}

KWY's Algorithm~2~\cite[lines~6 and~8]{KohWeinsteinYing} uses
target $(1+\xi)\lambda$ instead of $U$. The following argument
allows any $U$ and needs no minimum-cut lower bound.

\begin{proof}
Initialization adds at most $\xi/b$ per column to a row, so
$\|A\alpha\|_\infty\le\xi$. Before stopping, $Ax\le\cng$ and
$\|\cng\|_\infty<\Gamma$. Each later update therefore multiplies
every active coordinate by at least $1+\xi/\Gamma$.
Before any noninitial update, an active column has survived all
$k$ preceding noninitial updates. Thus
$\alpha_S(1+\xi/\Gamma)^k\le x_S<\Gamma/\max_e A_{e,S}$.
Substituting $\alpha_S$ gives $(1+\xi/\Gamma)^k<b\Gamma/\xi$.
Using $\log(1+\xi/\Gamma)\ge\xi/(2\Gamma)$ yields
$k<(2\Gamma/\xi)\log(b\Gamma/\xi)$. Initialization and the final
update are included in $J_b$; a call with no later update is immediate.
\end{proof}

\subsection{Building a sparsifier in one pass}
For each cut search, we build a small reweighted graph that
approximately preserves the current cut values. The update history
stays fixed throughout this pass.

\begin{theorem}[One-pass cut sparsification]\label{thm:sparsifier}
Fix a weighted insertion-only stream on $n$ vertices with at most
$M$ edges. For $0<\zeta,\upsilon<1/2$,
$\StreamSparsify\allowbreak(\zeta,\upsilon)$ makes one pass and
returns \textnormal{\textsc{Fail}} or a reweighted subgraph $H$.
With probability at least $1-\upsilon$, it returns $H$ satisfying
$(1-\zeta)q(\delta(S))\le q_H(\delta(S))\le(1+\zeta)q(\delta(S))$
for every cut $S$.
On every run, it uses $\wtO(n\zeta^{-2})$ words and
$\wtO((M+n)\zeta^{-2})$ arithmetic operations, including all
intermediate computations and output processing. A returned $H$
has $\wtO(n\zeta^{-2})$ edges with their original identifiers.
Disconnected graphs are allowed. Here $\wtO$ also hides logarithms
in $1/\upsilon$.
\end{theorem}

Maintain summaries of small edge blocks. When two summaries represent
equally many blocks, sparsify their union using
Kyng--Pachocki--Peng--Sachdeva's (KPPS) Theorem~1.2~\cite{KyngPachockiPengSachdeva}.
Each block undergoes only logarithmically many reductions. A smaller
error per reduction bounds the total error by $\zeta$; the final
summaries together form $H$. Lemma~\ref{lem:word-sparsifier} gives
the construction and its integer implementation.

\section{Compressing and reconstructing updates}\label{sec:replay}

We first show that repeated updates have a short record for any fixed
nonnegative matrix. We then use tree-path queries to evaluate that
record for cuts, without storing a weight for every edge.

\subsection{The update record}
Use the rule of \cref{prop:focus-interface}, starting with $w_e^{(0)}=1$ and $\cng_e^{(0)}=0$. If $h_e^{(t)}:=(Ag^{(t)})_e$ is the load added to row $e$, then
\begin{equation}\label{eq:weight-update}
 w_e^{(t+1)}=w_e^{(t)}(1+h_e^{(t)}),\qquad
 \cng_e^{(t+1)}=\cng_e^{(t)}+h_e^{(t)}.
\end{equation}
Each update adds at most $\xi$ per row; after initialization, the largest increase equals $\xi$.

\begin{lemma}[Identical updates within a phase]\label{lem:phase-compression}
Under the rule of \cref{prop:focus-interface}, the active coordinates of
$x$ remain proportional to their initial values $\alpha_S$. For a fixed
active set $\mathcal A$, every update after initialization is therefore
$g=\sigma\alpha|_{\mathcal A}$, where
$\sigma:=\xi/\|A(\alpha|_{\mathcal A})\|_\infty$.
\end{lemma}
\begin{proof}
Initially $x=\alpha$. If $x_S=p\alpha_S$ on the active set, the common
factor cancels: $g=\xi p\alpha|_{\mathcal A}/(p\|A(\alpha|_{\mathcal A})\|_\infty)
=\sigma\alpha|_{\mathcal A}$. Each surviving coordinate becomes
$(p+\sigma)\alpha_S$, so proportionality is preserved. Since $A\ge0$
and the row weights only increase, a removed column never becomes active again.
\end{proof}

A \emph{phase} is a maximal nonempty sequence of updates after
initialization with the same active columns; it may require several passes.
Let $\nph$ be the number of phases, and let $\sigma_p,k_p$ be the update
scalar and repetition count in phase $p$. For each column $S$, record its
last active phase $d(S)$, or zero if initialization removes it. If the call
is interrupted, give surviving columns $d(S)=\nph$, also when $\nph=0$.
For $1\le p\le\nph$, column $S$ is active exactly when $d(S)\ge p$.

\Needspace{11\baselineskip}
\begin{theorem}[Compressed update record]\label{thm:focus-record}
A call of \cref{prop:focus-interface} that initializes a group $B$ of
$b\ge1$ columns has at most $b$ phases, even if the stopping rule interrupts it.
Given these columns, its history needs only $(\alpha_S,d(S))$ per column
and $(\sigma_p,k_p)$ per phase: $O(b)$ numbers and indices.
This excludes storage or access to the matrix and its columns.

For row $e$, let $h_e^0:=\sum_{S\in B}A_{e,S}\alpha_S$ be the
initial load and let
$h_e^p:=\sigma_p\sum_{S\in B:\,d(S)\ge p}A_{e,S}\alpha_S$
be one update's load in phase $p$, for $1\le p\le\nph$.
The complete call has the exact effect
\begin{equation}\label{eq:compressed-effect}
 \begin{aligned}
 w_e&\leftarrow w_e(1+h_e^0)\prod_{p=1}^{\nph}(1+h_e^p)^{k_p},\\
 \cng_e&\leftarrow\cng_e+h_e^0+\sum_{p=1}^{\nph}k_ph_e^p.
 \end{aligned}
\end{equation}
\end{theorem}
\begin{proof}
At least one column leaves between consecutive phases, so there are at
most $b$ phases. The indices $d(S)$ recover every active set without a
separate list. By \cref{lem:phase-compression}, phase $p$ repeats one
update $k_p$ times. Equation~\eqref{eq:weight-update} therefore gives the
multiplier $(1+h_e^p)^{k_p}$ and added load $k_ph_e^p$.
Combining the phases with the initial update gives
\eqref{eq:compressed-effect}. A call returning before initialization
needs no record.
\end{proof}

For Held--Karp, the record also stores each cut's one- or two-edge
description and a pointer to its shared $O(n)$-word tree. The queries
below evaluate the required row sums; Appendix~\ref{app:precision}
bounds the record by $\wtO(b)$ words.

The repetition counts are essential: row increments $a,b$ multiply a
weight by $(1+a)(1+b)$, not $1+a+b$. Compression saves storage, but does
not find new groups or determine when a phase ends. Those tasks still
require cut searches and stream passes.

To recover the best estimate, save a \emph{checkpoint}: the initial
state or a state just after an update or an increase of $\lambda$.
Write $w^{(t)}$, $D_t:=\sum_e w_e^{(t)}$, and $\lambda_t$ for these
quantities at checkpoint $t$. Save the one minimizing $D_t/\lambda_t$,
including its own lower bound $\lambda_t$.

\begin{lemma}[Recovering a saved cover]\label{lem:snapshot}
The history and $O(1)$ saved indices and numbers recover the selected
weights $w^*$ and bound $\lambda_*$. If
$0<\lambda_*\le\min_{S\in\mathcal C}\sum_{e\in\delta(S)}w_e^*/c_e$,
then $y_e^*:=2w_e^*/(\lambda_*c_e)$ is a feasible demand-two cover
of cost $2\sum_e w_e^*/\lambda_*$.
\end{lemma}
Initialization and \cref{thm:clear-epoch} establish this hypothesis on $\Good$ (Section~\ref{sec:assembly}).

\begin{proof}
Whenever the ratio improves, save $\lambda_t$ and the current call and
update indices. After the call ends, express that update index as a phase
and repetition count, also recording whether initialization is included.
Evaluate earlier calls in full and the selected call only up to this point.
A change in $\lambda$ changes the saved bound, not the weight history.

This reconstructs the selected state on any run. When its bound is valid,
$y^*(\delta(S))=2\sum_{e\in\delta(S)}w_e^*/(\lambda_*c_e)\ge2$,
proving feasibility. With rounding, we select an upper estimate of the
cover's cost instead; the same conclusion holds on the success event.
\end{proof}

\subsection{Using the record in stream passes}
For cuts, $A_{e,S}=1/c_e$ when $e$ crosses $S$, and zero otherwise.
We need two operations: sum the coefficients of cuts crossed by an edge,
and add the edge's capacity to those cuts. Both use the path $P_T(u,v)$
between its endpoints in the reference tree.

\begin{lemma}\label{lem:path-parity}
For a cut $S$ represented by at most two edges of $T$, an input edge $uv$ crosses $S$ exactly when $|\delta_T(S)\cap P_T(u,v)|$ is odd.
\end{lemma}
\begin{proof}
Along the tree path, crossing a boundary edge switches sides of $S$. Its endpoints lie on different sides exactly when the number of switches is odd.
\end{proof}

\begin{proposition}[Tree-path queries]\label{prop:range-oracle}
For $b$ distinct cuts crossing at most two edges of $T$, store their
tree boundaries, coefficients $\alpha_S$, and indices
$d(S)\in\{0,\ldots,\nph\}$. Using $\wtO(b)$ space and preprocessing,
we support the following operations:
\begin{enumerate}[(i)]
\item Given $uv$ and $0\le p\le\nph$, sum $\alpha_S$ over the cuts it crosses with $d(S)\ge p$. Phase zero includes all cuts.
\item Given $(uv,\omega)$, add $\omega$ to the running total for each cut
crossed by $uv$. Start the totals at zero and report them after the pass.
\item Instead of adding, keep the minimum $\omega$ seen for each crossed
cut, starting at $+\infty$.
\end{enumerate}
Range trees use $O(b\log^2(2+b))$ words and
$O(b\log^3(2+b))$ preprocessing. Each operation costs
$O(\log^5(2+n))$ in (i) or $O(\log^4(2+n))$ in (ii)--(iii).
Reporting then costs $O(\log^2(2+b))$ per cut.
The shared tree uses another $O(n)$ words.
\end{proposition}
\begin{proof}
Number the tree edges so each decomposition path is consecutive.
Store a single-edge cut as $(\operatorname{pos}(a),d(S))$ and a two-edge
cut as $(\operatorname{pos}(a),\operatorname{pos}(b),d(S))$, with a fixed
order on the pair. The positions $\mathcal I$ on $P_T(u,v)$ and their
complement $\overline{\mathcal I}$ are each unions of $O(\log n)$
disjoint intervals. By \cref{lem:path-parity}, a query selects points in
\[
 \mathcal I\times[p,\nph]\quad\text{or}\quad
 (\mathcal I\times\overline{\mathcal I}\times[p,\nph])
 \cup(\overline{\mathcal I}\times\mathcal I\times[p,\nph]).
\]
For a pair, exactly one position must lie on the path. These sets are
unions of $O(\log^2 n)$ disjoint boxes, each a product of intervals.
Range sums over the boxes give (i).

For (ii), omit the phase coordinate and decompose each query rectangle
into disjoint stored subsets. Record the addition on each subset;
a cut's total is the sum of additions on subsets containing it.
Use minimum instead of addition for (iii)~\cite{deBergEtAl}.
Since $b\le(n-1)+\binom{n-1}{2}$, the usual range-tree bounds give the
stated costs. Duplicate cuts may be merged. Disjoint sums avoid
subtraction and cancellation when rounding.
\end{proof}

For a cut column, $\alpha_S=(\xi/b)\min_{e\in\delta(S)}c_e$.
A phase-zero query divided by $c_e$ gives $h_e^0$; a phase-$p$ query
multiplied by $\sigma_p/c_e$ gives $h_e^p$. Thus one completed record
takes $\wtO(\nph+1)$ work per edge, including initialization when
there are no later phases.

During an unfinished call, the final indices $d(S)$ are unknown.
Keep one query structure per update and compress them when the call ends.
This uses $\wtO(b\xi^{-2})$ temporary words by the $J_b$ bound.
Updates are computed from $\alpha$ and the active set; no copy of the
local vector $x$ is needed.

\begin{lemma}[Updating one group in the stream]\label{lem:streaming-focus}
Fix $b\ge1$ selected cuts, each described by one or two edges of the
same reference tree, with current values below $U$. Then $\Focus$
either raises them all to $U$ or reaches the load limit $\Gamma$. It uses at most
$J_b=\wtO(\xi^{-2})$ updates, $O(1)$ passes per update, and
$\wtO(b\xi^{-2})$ temporary words. Reconstructing an arriving edge's
weight costs $\wtO(J_b+1)$ plus $\wtO(1)$ per completed phase and
nonempty completed group. An empty group returns immediately.
\end{lemma}
\begin{proof}
One pass computes all initial coefficients using the minimum query.
For each later active set $\mathcal A$, a pass computes
\[
 \|A(\alpha|_{\mathcal A})\|_\infty
 =\max_{e=uv}\frac1{c_e}
        \sum_{\substack{S\in \mathcal A:\,|\delta_T(S)\cap P_T(u,v)|\text{ odd}}}\alpha_S,
\]
which determines $\sigma$. Record the update and make another pass to
recover edge weights, total weight, and maximum accumulated load.
Adding $q_e=w_e/c_e$ to each crossed cut gives all current cut values
in $B$. Remove cuts that reach $U$ and compare the new estimate with
the best one saved.

These operations implement \cref{prop:focus-interface}, so its iteration
bound applies. Each edge needs one query per completed phase, one per
completed group's initialization, and at most $J_b$ for the current group.
This gives the work bound.
\end{proof}
\FloatBarrier

\section{Raising the minimum-cut lower bound}\label{sec:buffer}
Assume $\Good$, so every randomized procedure meets its guarantee.
Section~\ref{sec:probability} bounds the failure probability.
A cut's \emph{true value} is $q(\delta(S))$ in the current input graph;
its \emph{stored value} is its weight in the sparsifier.

An outer round keeps $\lambda\le\mincut(q)$ fixed and aims to raise
every cut to $L:=(1+\xi)\lambda$, where $0<\xi\le1/20$.
Selected groups are raised to the larger target
$U:=(1+4\xi)\lambda$ to allow for search error. A completed round
sets $\lambda\gets L$; the global load limit may interrupt it.
We prove progress for one stage, then one reference tree, then all cuts.

\subsection{Search graphs and thresholds}
\begin{lemma}[Retaining the reference tree]\label{lem:T-preserving}
For $0<\zeta<1$, keeping a spanning tree $T$ with its exact capacities
and sparsifying $G-T$ to relative error $\zeta$ gives a
$(1\pm\zeta)$ cut sparsifier containing $T$.
\end{lemma}
\begin{proof}
For any cut, let $a\ge0$ be its tree contribution and $b\ge0$ the rest. Only $b$ changes, by at most $\zeta b\le\zeta(a+b)$.
\end{proof}

During a pass, use a dictionary of tree-edge identifiers to send only
non-tree edges to $\StreamSparsify$. Appendix~\ref{app:precision}
uses the same construction with rounded capacities.

For a reference tree with $r$ stages, put $\zeta:=\xi/[100(r+1)]$ and $\Lambda_{-1}:=U$. For $0\le j<r$, define
\begin{equation}\label{eq:stage-thresholds}
 \theta_j:=(1-\zeta)\Lambda_{j-1},\qquad
 \Lambda_j:=\frac{\theta_j}{1+\zeta}.
\end{equation}
The search threshold is $\theta_j$. A true value at least
$\Lambda_{j-1}$ is stored as at least $\theta_j$; a stored value at
least $\theta_j$ guarantees a true value at least $\Lambda_j$.
Thus $\Lambda_j$ is the lower bound preserved after stage $j$.
Table~\ref{tab:thresholds} distinguishes these bounds from $L$ and $U$.

\begin{table}[!htbp]
\caption{Search thresholds and cut-value lower bounds.}
\label{tab:thresholds}
\centering
\begin{tabularx}{\textwidth}{@{}p{1.5cm}p{3.0cm}X@{}}
\toprule
Threshold & Measured in & Purpose\\
\midrule
$L$ & input graph & target for ending the outer round\\
$U$ & input graph & target for each selected cut\\
$\theta_j$ & sparsifier $H_j$ & threshold for finding and checking cuts\\
$\Lambda_j$ & input graph & lower bound after stage $j$\\
\bottomrule
\end{tabularx}

\end{table}

\begin{lemma}[Threshold slack]\label{lem:threshold-slack}
The last lower bound satisfies $\Lambda_{r-1}>L$.
\end{lemma}
\begin{proof}
Each step multiplies the previous bound by $(1-\zeta)/(1+\zeta)\ge1-2\zeta$. Bernoulli's inequality gives $\Lambda_{r-1}\ge U(1-2\zeta r)\ge U(1-\xi/50)>(1+\xi)\lambda=L$.
\end{proof}

\subsection{From one stage to all cuts}
At stage $j$, build $H_j$, set $B_j:=\StaticExtract(H_j,T,\theta_j,j)$, and run $\Focus(B_j,U)$ unless the group is empty.

\begin{lemma}[Progress in one stage]\label{lem:stage-transfer}
On $\Good$, suppose every cut assigned before stage $j$ has true value
at least $\Lambda_{j-1}$; stage zero needs no such assumption.
Then the search in $H_j$ is valid and returns only cuts of true value
below $U$. Unless the load threshold stops the call, every cut in
$\mathcal F_j(T)$ has true value at least $\Lambda_j$ afterwards.
\end{lemma}
\begin{proof}
Earlier cuts have stored value at least
$(1-\zeta)\Lambda_{j-1}=\theta_j$, so \cref{prop:static-package}
applies; stage zero has none to check. For a cut returned by the search,
$q(\delta(S))\le q_{H_j}(\delta(S))/(1-\zeta)
<\theta_j/(1-\zeta)\le U$.
Unless the algorithm stops, $\Focus$ raises it to $U>\Lambda_j$.
An assigned cut not returned has stored value at least $\theta_j$,
so $q(\delta(S))\ge q_{H_j}(\delta(S))/(1+\zeta)\ge\Lambda_j$.
Later updates only increase these values.
\end{proof}

\begin{theorem}[Progress for one reference tree]\label{thm:clear-tree}
On $\Good$, the $r=O(\log n)$ stages either reach the load limit
$\Gamma$ or raise every cut crossing at most two edges of $T$ to
at least $L$. Each stage updates a group of $\wtO(n)$ cuts.
\end{theorem}
\begin{proof}
Induct on $j$. Earlier cuts remain above their established bounds
because capacities only increase. A cut assigned at stage $i<j$
therefore has value at least $\Lambda_i\ge\Lambda_{j-1}$,
since these bounds decrease with the stage number.
Apply \cref{lem:stage-transfer} at each stage, including the final
cross-path stage. All cuts represented by $T$ are assigned a stage,
and \cref{lem:threshold-slack} gives $\Lambda_{r-1}>L$.
\end{proof}

At the start of a round, build $H^{\rm start}$ at accuracy
$\zeta_0:=\xi/20$ and choose its reference trees using
\cref{thm:tree-packing}. Keep these trees for the round;
each stage builds its own sparsifier $H_j$.

\begin{lemma}[Coverage by the round-start trees]\label{lem:epoch-packing}
On $\Good$, any cut below $L$ during the outer round crosses at most two edges of a reference tree chosen at its start.
\end{lemma}
\begin{proof}
Such a cut was also below $L$ at the start, since capacities only
increase. The minimum cut then was at least $\lambda$, so in
$H^{\rm start}$ its ratio to the minimum is below
$\frac{1+\zeta_0}{1-\zeta_0}(1+\xi)<5/4$.
Theorem~\ref{thm:tree-packing} therefore represents it in one of the
chosen trees.
\end{proof}

\begin{theorem}[Progress in one outer round]\label{thm:clear-epoch}
On $\Good$, processing all reference trees of an outer round either
reaches the load limit $\Gamma$ or establishes $\mincut(q)\ge L$.
In the latter case, setting $\lambda\gets L$ preserves a valid lower bound.
\end{theorem}
\begin{proof}
Processed cut values do not decrease. A cut remaining below $L$ would be represented by a reference tree by \cref{lem:epoch-packing}, contradicting \cref{thm:clear-tree} for that tree.
\end{proof}
\FloatBarrier

\section{Algorithm and analysis}\label{sec:assembly}
Handle $n=1$ and $m=1$ directly; assume $n,m\ge2$ below.
The first pass stores an exact MST by cycle replacement, using a
Sleator--Tarjan dynamic-tree data structure~\cite{SleatorTarjan};
this takes $O(m\log n)$ time and $O(n)$ words. It also counts $m$
and finds $\ell_c$.
Choose $\xi=\Theta(\eps)$ with $\xi\le\eps/40$, and start with
$w_e=1$, $\cng_e=0$, and $D_0=m$. Sparsify the initial capacities
$q_e=1/c_e$ at accuracy $\zeta_0=\xi/20$. Compute the stored graph's
minimum cut using Karger's randomized near-linear-time
algorithm~\cite{Karger}, subject to the work limit and failure budget
specified below. Divide the resulting value by $1+\zeta_0$ to
obtain $\lambda_0$.

Keep $Z:=\min_t D_t/\lambda_t$ and its checkpoint as in
\cref{lem:snapshot}, starting with the initial state.
Each completed round increases $\lambda$. Stop when a row's
accumulated load reaches $\Gamma=\log m/\xi$.

Fix execution limits before drawing randomness, so failed runs also
obey the resource bounds. Choose a sufficiently large polynomial
$\CapsPoly$ in $\log n,\log m,\log(1/\xi)$, with coefficients
allowed to depend on $c_0$, and set
\begin{equation}\label{eq:deterministic-caps}
 \begin{aligned}
 E_{\max}&:=\lceil\CapsPoly\xi^{-2}\rceil,&
 B_{\max}&:=\lceil n\CapsPoly\xi^{-2}\rceil,\\
 I_{\max}&:=\lceil\CapsPoly\xi^{-4}\rceil,&
 N_{\max}&:=I_{\max}.
 \end{aligned}
\end{equation}
These limit outer rounds, total cut descriptions, updates, and
randomized calls, respectively. The polynomial absorbs the procedures'
logarithmic factors, including $\log N_{\max}$ for amplification.
The limits have no polynomial dependence on $\ell_c$: cost encoding
length affects word size (Section~\ref{sec:prelim}), not these counts.

Check limits before each operation or allocation. Also enforce the
per-group limit $J_b$ (up to a constant for rounding), the sparsifier
limits in \cref{thm:sparsifier}, and the amplified near-linear work and
storage limits for all stored-graph routines, including initial minimum
cut. Return \textnormal{\textsc{Fail}} before exceeding a limit, or on
a disconnected initial sparsifier, a nonpositive initial bound, or an
invalid reference tree.

Algorithm~\ref{alg:master} stores weights and loads implicitly in the
history; its assignments do not allocate edge-indexed vectors.
$\StaticExtract$ includes the cut-value checks, and $\Focus$ compares
$D/\lambda$ with $Z$ after every update. Make the same comparison
after increasing $\lambda$. On $\Good$, initialization and
\cref{thm:clear-epoch} keep $\lambda$ valid, so each ratio is a
feasible demand-one cover's cost.

\begin{algorithm}[!htb]
\caption{Semi-streaming Held--Karp estimator in exact arithmetic}\label{alg:master}
\begin{algorithmic}[1]
\State Make the first pass; set limits \eqref{eq:deterministic-caps} and initialize counters
\State Start an empty history (implicitly $w_e=1$, $\cng_e=0$); set $D_0\gets m$
\State Build the initial sparsifier; compute $\lambda_0$ by the minimum-cut routine above
\State Set $\lambda\gets\lambda_0$, $Z\gets D_0/\lambda_0$, and the saved indices
\While{$\|\cng\|_\infty<\Gamma$}
 \State Count a new outer round; set $L\gets(1+\xi)\lambda$, $U\gets(1+4\xi)\lambda$
 \State Build $H^{\rm start}$ with \StreamSparsify; compute its reference trees
 \For{each reference tree $T$}
  \State Let $r$ be its stage count; set $\zeta\gets\xi/[100(r+1)]$, $\Lambda_{-1}\gets U$
  \For{$j=0,\ldots,r-1$}
   \State Set $\theta_j$, $\Lambda_j$ by \eqref{eq:stage-thresholds}
   \State Build $H_j$ using \StreamSparsify on non-tree edges; retain $T$
   \State $B_j\gets\StaticExtract(H_j,T,\theta_j,j)$
   \If{$B_j\ne\varnothing$}
    \State Count $|B_j|$ cuts; run $\Focus(B_j,U)$ within both update limits
   \EndIf
   \If{$\|\cng\|_\infty\ge\Gamma$} \State Exit both for-loops \EndIf
  \EndFor
 \EndFor
 \If{$\|\cng\|_\infty<\Gamma$}
  \State $\lambda\gets L$; compare $D/\lambda$ with $Z$ and save this state if smaller
 \EndIf
\EndWhile
\State \Return $2Z$ and the selected history indices
\end{algorithmic}
\end{algorithm}
\FloatBarrier

\subsection{Accuracy}
Let $X:=\sum_t\one^\top g^{(t)}$ be the total packing weight added,
and $R:=\|A\sum_tg^{(t)}\|_\infty$ the maximum accumulated row load.
We compare two bounds on the growth of $D$: one from the selected
columns, the other from a row with load $R$. Scaling all updates by
$1/R$ gives a feasible packing. The parameter $\beta$ also allows
for rounding.

\begin{lemma}[Two-threshold MWU bound]\label{lem:MWU-bound}
Suppose $0<\lambda_t\le\min_S(A^\top w^{(t)})_S$ at every
checkpoint and update. If every updated column has value at most
$\beta\lambda_t$ for some $\beta>0$, each update adds at most $\xi$
per row, and $(1-\xi)R>\log m$, then
\[
 z\le Z\le
 \frac{\beta X}{(1-\xi)R-\log m}
 \le\frac{\beta R}{(1-\xi)R-\log m}\,z.
\]
In particular, $\beta=1+4\xi$ and $R\ge\Gamma=\log m/\xi$ give $z\le Z\le(1+4\xi)z/(1-2\xi)$.
\end{lemma}
\begin{proof}
Every $w^{(t)}/\lambda_t$ is a feasible normalized cover, so
$D_t/\lambda_t\ge z$ and $Z\ge z$. Let $D_t,\lambda_t$ be the
values just before update $t$. These are checkpoints, hence
$\lambda_t/D_t\le1/Z$.

Only columns of value at most $\beta\lambda_t$ are updated. Thus
\[
 D_{t+1}=D_t+\langle A^\top w^{(t)},g^{(t)}\rangle
 \le D_t+\beta\lambda_t\one^\top g^{(t)}
 \le D_t\exp\!\left(\frac{\beta\one^\top g^{(t)}}{Z}\right).
\]
Multiplying and using $D_0=m$ gives $D_{\rm final}\le m\exp(\beta X/Z)$.
For the lower bound, choose a row with total load $R$.
Since $\log(1+u)\ge(1-\xi)u$ for $0\le u\le\xi$,
$D_{\rm final}\ge\prod_t(1+(Ag^{(t)})_e)\ge e^{(1-\xi)R}$.
Taking logarithms gives $((1-\xi)R-\log m)Z\le\beta X$.
The packing $\sum_tg^{(t)}/R$ is feasible, so $X\le Rz$.
When $R\ge\log m/\xi$, the denominator is at least $(1-2\xi)R$,
which gives the final bound.
\end{proof}

\subsection{Passes, memory, and running time}\label{sec:probability}
First assume $\Good$. Initially $q_e^{(0)}=1/c_e$.
Assign packing weight $1/\mincut(q^{(0)})$ to a minimum cut and
zero to the others. This is feasible because each crossing edge has
$1/c_e\le\mincut(q^{(0)})$. Hence $\mincut(q^{(0)})\ge1/z$ and
$\lambda_0\ge1/((1+O(\xi))z)$.

Before stopping, $w_e\le e^{\cng_e}<m^{1/\xi}$, so
$D<m^{1+1/\xi}$. Since $D/\lambda\ge z$, we have
$\lambda<m^{1+1/\xi}/z$. After $e$ completed outer rounds,
$\lambda\ge(1+\xi)^e\lambda_0$, giving
$e\log(1+\xi)<(1+1/\xi)\log m+O(\xi)$.
Thus there are $O(\log m/\xi^2)$ completed rounds and at most one
interrupted final round. The bound also holds when each completed
round increases $\lambda$ by at least $1+\xi/2$, as after rounding.

Let $F$ count group-update calls, $B_f$ be the group in call $f$, and $I$ count updates. Each outer round has polylogarithmically many trees and stages. The candidate-list bound and \cref{lem:streaming-focus} give
\begin{equation}\label{eq:resource-counts}
 F=\wtO(\xi^{-2}),\qquad
 \sum_{f=1}^F|B_f|=\wtO(n\xi^{-2}),\qquad
 I=\wtO(\xi^{-4}).
\end{equation}
Only polylogarithms in $n,m,1/\xi$ occur in these counts, so the
caps in \eqref{eq:deterministic-caps} suffice without an $\ell_c$
factor. The total number of noninitial phases is at most $I$.

Including empty searches, there are $\wtO(\xi^{-2})$ stages. Each
nonempty group uses $\wtO(\xi^{-2})$ updates and $O(1)$ passes per
update, for $\wtO(\xi^{-4})$ passes overall. The two factors come
from outer rounds and updates per group; trees and stages per round
contribute only logarithmic factors.

\begin{lemma}[Termination within the execution limits]\label{lem:caps}
The algorithm terminates within its resource limits on every execution. On $\Good$, it reaches the load limit $\Gamma$ without exceeding a limit.
\end{lemma}
\begin{proof}
An operation that would exceed a limit returns \textnormal{\textsc{Fail}}. On $\Good$, the bounds on outer rounds, updates, and individual groups lie below the limits when their constants are large enough. If the maximum load never reached $\Gamma$, the finite group calls would complete enough outer rounds to force $\lambda\ge m^{1+1/\xi}/z$, contradicting the upper bound above. Thus a successful run stops at the load limit.
\end{proof}

By \eqref{eq:resource-counts}, completed records use
$\sum_f\wtO(|B_f|)=\wtO(n\xi^{-2})$ words. Their reference trees
must remain stored: newly sampled trees would not reproduce the recorded
cuts. Including repetitions to reduce failure probability, there are
$\wtO(\xi^{-2})$ trees, each using $O(n)$ words.
Discard a tree only when no record refers to it.

The current group and sparsifier each use $\wtO(n\xi^{-2})$ words by \cref{lem:streaming-focus,thm:sparsifier}; the MST and dictionaries need $O(n)$. The execution limits enforce these bounds on every run. Table~\ref{tab:resources} in Appendix~\ref{app:precision} collects the individual terms.

For each edge, evaluate at most $I$ completed phases, $F$ initial
updates, and $J_b$ updates in the current group. Proposition~\ref{prop:range-oracle}
gives $O(m(I+F+J_b+1)(\log^5(2+n)+\log(2+I_{\max})))$ work per
pass for this reconstruction. The final logarithm accounts for computing
powers by repeated squaring; $F$ includes calls ending at initialization.
This counts reconstruction work, not the worst-case work at one edge
arrival during sparsifier construction.

Building a sparsifier costs
$\wtO((m+n)\zeta^{-2})=\wtO(m\xi^{-2})$, since $n\le m+1$ and
$\zeta^{-2}=\wtO(\xi^{-2})$. Retaining tree edges, merging identifiers,
and rounding add $\wtO(m+n\zeta^{-2})$ work. Minimum cut~\cite{Karger},
reference-tree construction, cut finding, and verification take
near-linear work on the $\wtO(n\zeta^{-2})$ stored edges.
Query preprocessing and reporting cost at most $\wtO(n\xi^{-4})$.

Together with history reconstruction, this gives
$\wtO(m\xi^{-4})$ work per pass, including processing after the
scan, and $\wtO(m\xi^{-8})$ work overall.

\emph{Success probability.}
Give each randomized call failure budget $n^{-c_0-2}/N_{\max}$, including the initial minimum-cut call and all internal sampling, size, and work events of $\StreamSparsify$. Conditional on the earlier history, its input is fixed before its fresh seed, so its guarantee applies. A union bound over the deterministic limit $N_{\max}$ gives $\Pr[\Good]\ge1-n^{-c_0-2}\ge1-n^{-c_0}$, without assuming earlier calls succeeded.

To reduce failure probability to $\upsilon$, take the median of
$O(\log(1/\upsilon))$ independent runs. Running them together shares
passes but multiplies space and work; running them in sequence instead
multiplies passes. Appendix~\ref{sec:confidence} proves the value guarantee,
not feasibility of the selected run's cover.

\begin{proof}[Proof of \cref{thm:main-HK}]
On $\Good$, \cref{thm:clear-epoch} keeps $\lambda$ valid, and \cref{lem:caps} gives termination. Lemma~\ref{lem:MWU-bound} bounds the exact-arithmetic estimate, and \cref{lem:snapshot} recovers its cover. Proposition~\ref{prop:finite-precision} shows that rounding gives an output at most $(1+10\xi)\HK(G)$ with the same resource bounds. Choose $\xi=\Theta(\eps)$ with $\xi\le\eps/40$ and a power-of-two
denominator, and use $\HK(G)=2z$. The probability calculation completes the proof.
\end{proof}
\FloatBarrier

\subsection{Limitations and open questions}\label{sec:limits}
Proposition~\ref{prop:one-pass-lb} rules out one pass on weighted graphs.
It does not show that our pass count or multi-pass space bound is optimal,
or that the required pass count grows with $n$.
Fewer outer rounds or batching repeated updates may improve the upper
bounds. Batching requires finding when the first active cut reaches its
target without one pass per repetition. Subdividing weighted edges does
not give an unweighted lower bound: the added vertices change both the
cut constraints and the input size.

Tour construction and a distance-query version remain open for this method.
Certifying the saved cover would require a lower bound on its minimum
cut; a returned cut supplies only an upper bound.
Mukhopadhyay--Nanongkai's randomized algorithm~\cite[Theorem~1.3]{MukhopadhyayNanongkai}
does not certify feasibility on every run. A deterministic minimum-cut
computation with safe rounding could certify feasibility, but not the
approximation ratio of a failed run. We do not perform this check.

\emph{Use of AI tools.}
Generative AI tools assisted proof exploration, exposition, source checks,
and the preparation of finite verification scripts. These scripts do not
implement the imported KWY or KPPS algorithms. The authors are responsible
for the mathematics, references, and submitted text.

\FloatBarrier
\clearpage
\bibliographystyle{plainurl}
\bibliography{references}
\clearpage
\appendix

\section{Details of the cut searches}\label{app:static-contract}
Fix a stored positive-weight graph $H$, a rooted tree $T$, and a search
threshold $\theta>0$. We explain how KWY's results give
\cref{prop:static-package}. Table~\ref{tab:kwy-import} lists the results
from the cited arXiv version~\cite{KohWeinsteinYing}.

\subsection{Pairs on the same path}
KWY's Definition~4.15~\cite{KohWeinsteinYing} gives the interval
splitting in \cref{def:stage-families}. Each single-edge cut and same-path
pair is assigned once. Intervals of one height are disjoint and have total
size $O(n)$ over all paths. Their logarithmic height gives $O(\log n)$ stages.

Use the middle shore and separator of \cref{lem:root-prerequisites}, as in
KWY's Definition~4.16~\cite{KohWeinsteinYing}. A nonroot pair lies within
one child interval and belongs to an earlier stage. Its value is unchanged
by contraction, so the earlier cut bounds supply the hypothesis of
Lemma~4.9~\cite{KohWeinsteinYing}. No bound on the current pairs is assumed.

That lemma lists all cuts below $\theta$ crossing at most two edges of the
contracted path, with list size linear in the path length. It therefore
includes every small pair assigned to the interval. The stage and weight
checks in $\StaticExtract$ retain these pairs. Summing over the intervals
at one height gives $\wtO(n)$ cuts to check. Stage zero checks single-edge
cuts directly.

\subsection{Pairs on different paths}
The final stage handles two-edge cuts whose edges lie on different
decomposition paths. The current weights identify which path pairs
need to be searched.

For disjoint vertex sets $X,Y$, let $q_H(X,Y)$ be the total weight
of edges between them. There are two cases.

\emph{Disjoint descendant subtrees.} If neither tree edge $e,f$ lies
below the other, then
\[
 q_H(\delta(V_e\cup V_f))
 =q_H(\delta(V_e))+q_H(\delta(V_f))-2q_H(V_e,V_f).
\]
Both singleton cuts have value at least $\theta$. If the pair cut is
below $\theta$, the identity gives
\[
 2q_H(V_e,V_f)>q_H(\delta(V_e)),\qquad
 2q_H(V_e,V_f)>q_H(\delta(V_f)).
\]
These are KWY's mutual cross-interest conditions
(Definition~4.19~\cite{KohWeinsteinYing}).

\emph{Nested descendant subtrees.} If $f$ lies below $e$, use
\[
 q_H(\delta(V_e\setminus V_f))
 =q_H(\delta(V_e))+q_H(\delta(V_f))-2q_H(V_f,V\setminus V_e).
\]
The same singleton bounds imply
\[
 2q_H(V_f,V\setminus V_e)
 >\max\{q_H(\delta(V_e)),q_H(\delta(V_f))\}
\]
for every pair below $\theta$. This implies KWY's down-interest condition
(Definition~4.20~\cite{KohWeinsteinYing}); reverse nesting is symmetric.

KWY's Definitions~4.22--4.23~\cite{KohWeinsteinYing} and
Lemma~4.24~\cite{KohWeinsteinYing} select path pairs and edges on each
path using these conditions. Every two-edge cut below $\theta$ appears
in a selected pair. Counting retained edges over all selected pairs gives
$O(n\log n)$. Recompute these sets in the current graph $H$.

For each selected pair, KWY's Algorithm~5~\cite{KohWeinsteinYing}
contracts the other tree edges to a path with a separator between the
two retained sets. A middle shore that excludes the separator uses two
edges from one retained set, hence from one original path. Its cut was
assigned earlier and already has value at least $\theta$.
Lemma~4.9~\cite{KohWeinsteinYing} therefore finds every small pair
in the contracted path. Its linear list-size bound sums to $O(n\log n)$.
This is the cross-path step in the proof of KWY's
Lemma~4.8~\cite{KohWeinsteinYing}.

\subsection{Candidate checks and running time}
The searches list every required cut. Filtering by stage and checking
weights therefore gives exactly the set in \cref{prop:static-package}.

\emph{Why the threshold is arbitrary.} For any $\theta>0$, set KWY's
accuracy to $1/4$ and its parameter $\lambda$ to $4\theta/5$, so
$(1+1/4)\lambda=\theta$. Lemma~4.9~\cite{KohWeinsteinYing} requires
only that nonroot cuts have value at least $\theta$; this source
parameter need not be a global minimum-cut lower bound. The path-pair
inequalities and size bounds remain valid.

\emph{Work and storage.} Build \textsc{CutValue} once on $H$ using KWY's Lemmas~4.7
and~4.25~\cite{KohWeinsteinYing}. The path searches in
Lemma~4.9~\cite{KohWeinsteinYing} take near-linear work in their total
path lengths; selecting path pairs in Lemma~4.24~\cite{KohWeinsteinYing}
takes near-linear work in $|E(H)|+n$. The searches use the stored
cut-query structure rather than storing each contracted graph in full.
The preceding size sums give the work and storage bounds.

\begin{table}[!htbp]
\caption{KWY results used in this paper, numbered as in the cited full version. Locators refer to~\cite{KohWeinsteinYing}; the iteration bound is proved here.}
\label{tab:kwy-import}
\centering
\begin{tabularx}{\textwidth}{@{}>{\raggedright\arraybackslash}p{5.0cm}X@{}}
\toprule
Result in KWY, arXiv v2 & Guarantee or construction used here\\
\midrule
Theorem 2.3 & Near-minimum cuts are represented by the sampled reference trees; near-linear construction work.\\
Lemma 4.7 (\textsc{CutValue} only); Lemma 4.25; Section 4.6 preprocessing & Static cut-value queries on the stored graph $H$, with near-linear preprocessing. We do not retain its evolving \textsc{Focus} structure on the original stream.\\
Section 4.3 before Lemma 4.9; Definitions 4.15--4.16; Algorithm 4 & Middle shores, separator vertices, and the fixed interval-stage construction.\\
Lemma 4.9, items 1--2 and work bound & Linear candidate lists containing every small path cut once nonroot cuts meet the same threshold.\\
Property 4.17; Definitions 4.19--4.23; Lemma 4.24 & The path-decomposition property, interested-pair enumeration, and $O(n\log n)$ total retained tree edges.\\
Algorithm 5, lines 4--9; proof of Lemma 4.8 in Section 4.5 & Reduction of every remaining small cross-path cut to a path search whose other pairs were processed earlier.\\
Algorithm 2, lines 6 and 8; Lemma 3.6 & Initialization and the common-scalar update. Proposition~\ref{prop:focus-interface} proves the iteration bound for our target $U$.\\
\bottomrule
\end{tabularx}

\end{table}
\FloatBarrier

\subsection{Why a sparsifier does not replace the LP}
\begin{remark}[]\label{rem:no-static-coreset}
On the unit-cost complete graph $K_n$, $n\ge2$, summing singleton
constraints in \eqref{eq:unit-cover} gives $\sum_e y_e\ge n/2$.
The feasible assignment $y_e=1/(n-1)$ attains this bound, so $z=n/2$.
Fix $0<\zeta<1/2$ with $\zeta^2n\ge c_{\rm samp}\log n$ for a
sufficiently large absolute constant $c_{\rm samp}$.
Keep each edge independently with probability
$\pi:=c_{\rm samp}\log n/(\zeta^2n)$ and give it weight $1/\pi$.
With high probability, the resulting $H$ is a $(1\pm\zeta)$ cut
sparsifier with $\wtO(n/\zeta^2)$ edges~\cite{KargerSampling}.
It is connected on this event because every nontrivial cut stays positive.
Let $z_H$ be the demand-one covering optimum when these weights
are used as costs. Singleton constraints give $z_H\ge n/(2\pi)$,
hence $z_H/z\ge1/\pi=\Omega(\zeta^2n/\log n)$.
Thus preserving cut values does not justify using sparsifier
weights as LP costs. This does not rule out other sparse LP
representations. Our sparsifiers only find small cuts under
$q_e=w_e/c_e$.
\end{remark}

Chekuri--Quanrud's Theorem~1.4~\cite{ChekuriQuanrudSparsification}
sparsifies a \emph{given} feasible demand-two cover to
$O(n\log n/\eps^2)$ edges with at most a $1+\eps$ cost increase,
with high probability. This changes fractional multiplicities,
not input costs, so it does not contradict the example.
The cover must be supplied first; the result does not by itself
compute one from the input stream.

\section{Finite-precision implementation}\label{app:precision}
Exact fractions can become too long to store. We round update
coefficients and evaluate their products at bounded precision.
The resulting sequence may differ from the ideal one, so we analyze
it directly: its record stays short, its cut bounds remain valid,
and its estimate has the required accuracy. Constants are not optimized.

\subsection{Storing and evaluating rounded updates}
Choose $\tau:=\xi/[10^6(\lceil\log_2(n-1)\rceil+3)]$. The stage bound in \cref{def:stage-families} gives $\tau\le\zeta/100$ for every $\zeta=\xi/[100(r+1)]$ and for $\zeta_0=\xi/20$.

Store numbers in binary floating-point form, with separate exponents.
Evaluate each positive quantity $v$ with a lower bound $v^-$ and an upper
bound $v^+$ satisfying
\begin{equation}\label{eq:interval-enclosure}
 v^-\le v\le v^+\le(1+\tau)v^-;
\end{equation}
represent zero exactly. Limit each operation's relative error to
$\tau/[10^8(m+n+I_{\max}+2)^{16}]$. The polynomial covers positive range
sums, totals within a pass, and repetition counts up to $I_{\max}$.
When computing powers by repeated squaring, account for error in all
$I_{\max}$ possible original factors, not just the squaring steps.
Combining the errors from sums, products, divisions, and maxima gives
\eqref{eq:interval-enclosure}.

Range queries sum disjoint subsets, avoiding subtraction of large,
close numbers. Reconstruct each weight from the stored coefficients,
not an earlier rounded weight. Its upper bound then depends only on
the edge and fixed history, not on the new sparsifier's random choices.

The significant part of each number needs
$O(\log(m+n+I_{\max}+2)+\log(1/\tau))$ bits; input quantities use
their $\ell_c$-bit encodings. Even on a failed run,
$\log w_e\le I_{\max}\xi$, so exponents need only
$O(\log(I_{\max}+\ell_c+\log m))$ bits.
Each stored number therefore uses $\wtO(1)$ input words.

\emph{One stored update per phase.}
Round each initial coefficient downward once, with
$\alpha_S/(1+\tau)\le\bar\alpha_S\le\alpha_S$, and initialize using
$g=\bar\alpha$. Whenever the nonempty active set changes to $\mathcal A$,
compute bounds on $v:=\|A(\bar\alpha|_{\mathcal A})\|_\infty$.
Choose a stored scalar with $\xi/((1+\tau)v^+)\le\bar\sigma\le\xi/v^+$
and reuse it until the active set changes.
Thus each update is exactly $g=\bar\sigma\bar\alpha|_{\mathcal A}$
for the stored coefficients, and
$\xi/(1+\tau)^2\le\|Ag\|_\infty\le\xi$.
Initialization adds at most $\xi$ per row; each later update has largest
row increase at least $\xi/2$. Active coordinates remain proportional
to $\bar\alpha$, so \eqref{eq:compressed-effect} still holds exactly.

Remove a cut only when its computed lower bound satisfies
$q(\delta(S))^-\ge U$. It then has true value at least $U$.
A remaining cut has true value below $(1+\tau)U$.
Since capacities increase, removed cuts never return.

Let $R:=\|\cng\|_\infty$, with computed upper bound $R^+$.
Store $\bar\Gamma\in[\Gamma,(1+\tau)\Gamma]$ and stop when
$R^+\ge\bar\Gamma$. Before an update, $R<(1+\tau)\Gamma$;
at termination, $R\ge\Gamma/(1+\tau)$.
Each surviving coordinate grows by at least
$1+\xi/[2(1+\tau)\Gamma]$ per later update.
The proof of \cref{prop:focus-interface} then bounds the updates by a
constant multiple of $J_b$, even if sparsification fails.

\subsection{One-pass sparsification}\label{sec:word-sparsifier}
We sparsify unions of stored edge blocks. Merging only summaries of
equally many blocks limits each edge to logarithmically many reductions.
A smaller error per reduction controls the total cut error.

\emph{The static sparsifier.}
We use the combinatorial routine in KPPS Theorems~1.2 and~4.1,
implemented in Figures~3--5~\cite{KyngPachockiPengSachdeva}.
For $0<\zeta_{\rm loc}<1/2$, it reduces an explicit $n$-vertex,
$m_H$-edge weighted graph to $\wtO(n\zeta_{\rm loc}^{-2})$ edges
in $\wtO((m_H+n)\zeta_{\rm loc}^{-2})$ work. Increasing the
logarithmic sampling and spanner parameters gives failure probability
at most $\upsilon_{\rm loc}\in(0,1/2)$; the bounds also hide
$\log(1/\upsilon_{\rm loc})$. No numerical linear-system solver is used.

The procedure stores the current graph and marked edges. It reuses
$\wtO(m_H+n)$ workspace to build each spanner, a sparse graph that
approximately preserves path distances. Let $\alpha\ge1$ be its oversampling parameter.
With parameter $4\alpha$, Figure~5~\cite{KyngPachockiPengSachdeva}
assigns estimates $1$ to marked edges and $1/(4\alpha)$ to the others.
Figure~3~\cite{KyngPachockiPengSachdeva} therefore keeps marked edges
surely and others with probability $1/4$, multiplying sampled weights
by four. Edge identifiers remain unchanged.

Enforce limits on output size, work, and the number of reduction steps,
returning \textnormal{\textsc{Fail}} if needed.
The proof of KPPS's Theorem~4.1~\cite{KyngPachockiPengSachdeva}
shows that the edge count decreases geometrically, giving
$O(\log(2+m_H))$ steps with high probability. Fix that limit with a
sufficiently large constant and include all failures in
$\upsilon_{\rm loc}$. Return graphs already below the size target
unchanged. Process disconnected components separately and retain
isolated vertices.

\begin{lemma}[Integer-weight sparsification]\label{lem:word-sparsifier}
Let a stream have $n$ vertices and at most $M\ge1$ distinct edge
identifiers, with positive integer weights at most $K_{\rm in}$.
Parallel edges have different identifiers. The procedure in
\cref{thm:sparsifier} uses one pass, $\wtO(n\zeta^{-2})$ words,
and $\wtO((M+n)\zeta^{-2})$ operations.
Let $s_{\rm mr}=\wtO(n\zeta^{-2})$ be the summary-size limit below.
Each output weight is an integer using
$O(\log(2+K_{\rm in})+\allowbreak\log(2+M)\log(2+s_{\rm mr}))$ bits.
For $0<\zeta,\upsilon<1/2$, with probability at least $1-\upsilon$,
no limit is exceeded and every cut is preserved within $1\pm\zeta$.
\end{lemma}
\begin{proof}
\emph{Construction.}
Set $\ell_{\rm mr}:=1+\lceil\log_2(M+1)\rceil$,
$\zeta_{\rm loc}:=\zeta/(4\ell_{\rm mr})$, and
$\upsilon_{\rm loc}:=\upsilon/[4(M+1)]$.
Fix a common output-size limit $s_{\rm mr}$ using these parameters.
Since $\zeta_{\rm loc}^{-2}=O(\zeta^{-2}\log^2(2+M))$,
we have $s_{\rm mr}=\wtO(n\zeta^{-2})$.

Read blocks of at most $s_{\rm mr}$ edges, initially stored exactly at
level zero. Keep at most one summary per level. When two summaries
have the same level, sparsify their union and move it up one level,
as in a binary counter. Stored summaries represent disjoint input
blocks. Return their union and the final partial block; no final
sparsification is needed.

\emph{Cut values.}
A level-$j$ summary represents $2^j$ blocks. On successful static calls,
its cut values lie between $(1-\zeta_{\rm loc})^j$ and
$(1+\zeta_{\rm loc})^j$ times the original values in those blocks:
unions add cut values, and each reduction contributes one factor
$1\pm\zeta_{\rm loc}$. No block undergoes more than $\ell_{\rm mr}$
reductions, and
\[
 \begin{aligned}
 (1-\zeta_{\rm loc})^{\ell_{\rm mr}}
   &\ge1-\ell_{\rm mr}\zeta_{\rm loc}=1-\zeta/4,\\
 (1+\zeta_{\rm loc})^{\ell_{\rm mr}}
   &\le e^{\zeta/4}\le1+\zeta/2.
 \end{aligned}
\]
Thus each summary and their final union satisfy the $1\pm\zeta$
bounds. A partial block is exact.

\emph{Space and work.}
At most $\ell_{\rm mr}$ summaries are stored, each with at most
$s_{\rm mr}$ edges. A reduction processes at most $2s_{\rm mr}$ edges
and reuses its workspace, giving $\wtO(n\zeta^{-2})$ total space.

If $M<s_{\rm mr}$, return the stream exactly. Otherwise there are at
most $\lceil M/s_{\rm mr}\rceil$ blocks and fewer merges than blocks.
Each merge costs $\wtO(s_{\rm mr}\zeta_{\rm loc}^{-2})$, so total
work, including input and output, is $\wtO((M+n)\zeta^{-2})$.
Allocate memory as edges arrive; a short stream needs no full-sized buffer.

\emph{Probability and limits.}
There are at most $M$ static calls. Each input is fixed before its fresh seed is drawn, so each conditional failure probability is at most $\upsilon_{\rm loc}$. Their union has probability at most $M\upsilon_{\rm loc}<\upsilon/4$. Reserve the remaining budget for finite randomness, described below. The output-size, round, and work limits hold on every run, not only on successful ones.

\emph{Integer weights.}
An edge starts with an integer weight and is only multiplied by four. One static call has $O(\log(2+s_{\rm mr}))$ rounds under its limit. An edge participates in at most $\ell_{\rm mr}$ calls along the merging tree. These two counts give the stated bit-length bound on every run. Different input blocks have distinct identifiers, so their union does not duplicate an input edge.
\end{proof}

\emph{Generating randomness with finitely many bits.}
KPPS's Figure~4~\cite{KyngPachockiPengSachdeva} groups reciprocal
weights into ranges and uses unweighted graph searches and minimum
spanning forests. Process only nonempty ranges. Compare integer
weights exactly and reciprocals by cross multiplication; no electrical
resistance is computed numerically. Sampling with probability $1/4$
uses two fair bits.

The unweighted clustering assigns random exponential priorities.
Let $N_{\rm cmp}$ be a fixed polynomial bound on all priorities and
comparisons under the execution limits, and set
$h:=\lceil\log((N_{\rm cmp}+2)/\upsilon)\rceil$.
Cap priorities at $Ch$ and round to multiples of $2^{-Ch}$.
Exponential tails bound the probability of exceeding the cap.
For fixed input, comparisons involve two fresh priorities and an integer
distance. The priority difference has bounded density. A union bound
over priority pairs and integer offsets therefore bounds the probability
of changing a comparison by $\upsilon/4$ for sufficiently large $C$.

Sample each truncated discrete distribution within total-variation distance $\upsilon/(4N_{\rm cmp})$, using $O(h)$ random bits and polynomial-in-$h$ arithmetic. This adds at most $\upsilon/4$ to the failure probability. Outside these events, the decisions agree with the real-priority procedure. This finite-randomness argument is separate from the rounding of MWU updates.

A common positive scale can be removed before sparsification and restored afterwards. The integer lengths above use $\wtO(1)$ input words in our applications. Merging summaries adds logarithmic factors but preserves the claimed $\zeta^{-2}$ dependence in space and work.

\subsection{Rounding the search graph}
Rounding cut values separately could break the identities used by the
search. Instead, round edge capacities once and answer all cut queries
on that graph. Values above a fixed cap can be reduced because the
search only needs cuts below $U$.

\begin{lemma}[Rounded search graph]\label{lem:integer-surrogate}
Suppose $0<\lambda\le\mincut(q)$ and $U=(1+4\xi)\lambda$.
Let $0<\zeta<1/10$ and $0<\tau\le\zeta/100$. One stream pass produces $\wtO(n/\zeta^2)$ stored edges with weights that are integer multiples of a common scale. On successful sparsification, the resulting graph $H$ satisfies
\begin{equation}\label{eq:threshold-surrogate}
 (1-\zeta)\min\{q(\delta(S)),4U\}
 \le q_H(\delta(S))\le(1+\zeta)q(\delta(S))
\end{equation}
for every cut. A prescribed reference tree keeps its rounded capacities exactly. After removing the common scale, edge weights have $O(\log(m/\tau))$ bits.
\end{lemma}
\begin{proof}
Set $\Delta:=\tau\lambda/m$, $K:=\lceil4U/\Delta\rceil$, and
$q'_e:=\Delta\min\{K,\lceil q_e^+/\Delta\rceil\}$.
This rounds capacities up to multiples of $\Delta$ and caps them at
$K\Delta$. Let $f,f'$ be a cut's values before and after this change.
Without capping, $f'\ge f$; otherwise a capped edge contributes at
least $4U$. Rounding adds at most $\Delta$ per edge. Since $f\ge\lambda$,
\begin{equation}\label{eq:first-surrogate}
 \min\{f,4U\}\le f'\le(1+\tau)f+m\Delta\le(1+2\tau)f.
\end{equation}

Remove the common factor $\Delta$, keep the tree with weights $q'_e/\Delta$, and sparsify the other edges at accuracy $\zeta/4$. The input integers lie in $[1,K]$, so \cref{lem:word-sparsifier} applies with $M=m$ and $K_{\rm in}=K$. Restore the scale and add the tree. Let $f_1$ be its value on the cut under consideration. On success,
$(1-\zeta/4)f'\le f_1\le(1+\zeta/4)f'$.

The returned weights are integer multiples of $\Delta$. Merge any
duplicate identifiers and cap each integer weight at $K$.
Tree weights are unchanged, and no further rounding is needed.
Capping cannot increase a cut value. If it reduces one, that cut still
has an edge of weight $K\Delta\ge4U$. Thus
$\min\{f_1,4U\}\le f_H\le f_1$ for the final cut value $f_H$.
With \eqref{eq:first-surrogate}, this gives
$f_H\ge(1-\zeta/4)\min\{f,4U\}$ and
$f_H\le(1+\zeta/4)(1+2\tau)f$.
Since $\tau\le\zeta/100$, \eqref{eq:threshold-surrogate} follows.

Compare a capacity with the cap before converting it to an integer.
Since $K=O(m/\tau)$, no oversized integer is created. Edge weights and
cut sums have the stated bit length. Sparsification takes
$\wtO((m+n)\zeta^{-2})$ work; merging identifiers and capping the
output take $\wtO(n\zeta^{-2})$.
\end{proof}

All cut values and path-pair identities are exact in the stored
integer graph. Compare a cut's integer sum with $\theta_j/\Delta$.
The factor $\lambda$ cancels, leaving a rational depending only on
$\xi,\zeta,m$ and $j=O(\log n)$, which uses $\wtO(1)$ words.

Use the same construction at accuracy $\zeta_0$ for later round-start
graphs.

\emph{The initial graph.}
Initialization has no minimum-cut lower bound yet. Instead, set
$\Delta_0:=\tau2^{-\ell_c}/m$ and
$q'_e:=\Delta_0\lceil1/(c_e\Delta_0)\rceil$.
Every original nontrivial cut has capacity at least $2^{-\ell_c}$. The total rounding increase is at most $m\Delta_0=\tau2^{-\ell_c}$, giving relative error at most $\tau$. The integers $q'_e/\Delta_0$ have $O(\ell_c+\log(m/\tau))$ bits. Apply $\StreamSparsify$ at accuracy $\zeta_0/4$ after removing the scale; its integer output needs no second rounding. On success the combined error is at most $1\pm\zeta_0$.

By \cref{lem:word-sparsifier}, the stored integers have $O(\ell_c+\operatorname{polylog}(n,m,1/\xi,1/\upsilon))$ bits, or $\wtO(1)$ input words. Divide the stored graph's exact minimum cut by $1+\zeta_0$ and round down within factor $1+\tau$. This gives a valid initial lower bound with
$\lambda_0\ge\frac{1-\zeta_0}{(1+\zeta_0)(1+\tau)}\mincut(q^{(0)})\ge\mincut(q^{(0)})/(1+O(\xi))$.

\subsection{Why the cut tests remain valid}
Write $f:=q(\delta(S))$. The cap at $4U$ in
\eqref{eq:threshold-surrogate} does not affect the three uses of
cut values in \cref{lem:stage-transfer}.

\emph{Earlier cuts.} If $f\ge\Lambda_{j-1}$, then
$\Lambda_{j-1}\le U<4U$, so
$q_H(\delta(S))\ge(1-\zeta)\Lambda_{j-1}=\theta_j$.

\emph{Returned cuts.} If $q_H(\delta(S))<\theta_j\le(1-\zeta)U$,
then $f\ge4U$ would contradict \eqref{eq:threshold-surrogate}.
For $f<4U$, it gives $f<U$. Such a cut is removed from the active
group only when its true value reaches $U$.

\emph{Cuts not returned.} An assigned cut not in the checked list has
$q_H(\delta(S))\ge\theta_j$, so
$f\ge q_H(\delta(S))/(1+\zeta)\ge\Lambda_j$.
Thus the stage induction is unchanged. The error $\zeta$ already
includes reconstruction, rounding, and sparsification. Capping only
reduces cut values; no further error enters the threshold recurrence.

\emph{Round-start trees.} Every stored cut is at least
$(1-\zeta_0)\lambda$, while a true cut below $L$ has stored value
at most $(1+\zeta_0)L$. Its ratio to the stored minimum is therefore
below $\frac{1+\zeta_0}{1-\zeta_0}(1+\xi)<\frac54$, so the
reference trees still cover every required cut.

After a completed round, round $(1+\xi)\lambda$ down within factor
$1+\tau$. The lower bound stays valid and grows by at least
$1+\xi/2$, preserving the outer-round count.

\subsection{Final accuracy and resource bounds}
At each checkpoint, round $D_t/\lambda_t$ upward within factor $1+\tau$. Let $Z_*$ be the minimum exact ratio for the sequence defined by the stored coefficients, and $Z^+$ the minimum rounded upper estimate. On $\Good$, $z\le Z_*\le Z^+\le(1+\tau)Z_*$. The selected history indices recover the cover in \cref{lem:snapshot}, of cost at most $2Z^+$.

\begin{proposition}[Accuracy with rounding]\label{prop:finite-precision}
The implementation preserves the pass, space, and arithmetic-operation bounds of \cref{thm:main-HK}. On $\Good$, its output $\widehat H=2Z^+$ satisfies
\begin{equation*}
 \HK(G)\le\widehat H\le
 \frac{(1+\tau)^2(1+4\xi)}{1-2\xi-\xi\tau}\,\HK(G)
 \le(1+10\xi)\HK(G).
\end{equation*}
Choose $\xi=\Theta(\eps)$ with $\xi\le\eps/40$ and a power-of-two
denominator to obtain the requested accuracy.
\end{proposition}
\begin{proof}
Apply \cref{lem:MWU-bound} to the exact sequence defined by the rounded coefficients. Each update has width at most $\xi$ and uses columns below $\beta\lambda_t$, with $\beta=(1+\tau)(1+4\xi)$. The stopping rule gives $\log m\le\xi(1+\tau)R$. Hence $Z_*\le\frac{(1+\tau)(1+4\xi)}{1-2\xi-\xi\tau}\,z$.
Rounding the reported ratio adds the other factor $1+\tau$.
For $\tau\le\xi/100$ and $\xi\le1/20$, the numerator is at most $1+4.025\xi$ and the denominator is at least $1-2.001\xi>0$. Since $(1+10\xi)(1-2.001\xi)-(1+4.025\xi)=3.974\xi-20.01\xi^2>0$, their ratio is at most $1+10\xi$.

Each coefficient uses $\wtO(1)$ words, and each group still has at most
$b$ phases. The safe cut tests and row-increase bounds change the update
and outer-round counts only by constants. Lemma~\ref{lem:word-sparsifier}
uses $\wtO(n/\zeta^2)$ words for each rounded graph and workspace,
and $\wtO((m+n)\zeta^{-2})$ operations to construct it, as counted in
Section~\ref{sec:assembly}. History and tree storage are unchanged.
Static calculations use $\wtO(1)$-word numbers. The fixed call limit
and fresh randomness give the probability bound of
Section~\ref{sec:probability}.
\end{proof}

Table~\ref{tab:resources} summarizes the resource bounds, including the rounded implementation.

\begin{table}[!htbp]
\caption{Space and work bounds from the counts in \eqref{eq:resource-counts}.}
\label{tab:resources}
\centering
\begin{tabularx}{\textwidth}{@{}>{\raggedright\arraybackslash}p{3.2cm}>{\raggedright\arraybackslash}p{3.2cm}X@{}}
\toprule
Object or operation & Bound & Reason\\
\midrule
Completed records & $\wtO(n\xi^{-2})$ words & sum of $\wtO(|B_f|)$\\
Retained trees & $\wtO(n\xi^{-2})$ words & all trees referenced by records\\
Current cut group & $\wtO(n\xi^{-2})$ words & $J_b$ query structures on $b=\wtO(n)$ cuts\\
Sparsifier and workspace & $\wtO(n\xi^{-2})$ words & summaries; $\zeta^{-2}=\wtO(\xi^{-2})$\\
One sparsifier construction & $\wtO((m+n)\xi^{-2})$ operations & total work in \cref{thm:sparsifier}\\
Passes & $\wtO(\xi^{-4})$ & constant passes per update\\
Replay per edge & $\wtO(\xi^{-4})$ & phases, initial factors, and current updates\\
\bottomrule
\end{tabularx}

\end{table}

\subsection{Reducing the failure probability}\label{sec:confidence}
The inverse failure budget has logarithm $(c_0+2)\log n+\log N_{\max}$. A larger fixed $c_0$ can increase sampling and reference-tree repetitions, but not the powers of $1/\eps$, the load limit, or the sparsifier accuracy.

With $c_0=3$ and $n\ge2$, a run fails with probability at most $1/8$. For target failure probability $0<\upsilon<1/2$, take $k:=2\lceil2\log(1/\upsilon)\rceil+1$ independent runs and return their median, treating \textsc{Fail} as $+\infty$. A successful majority puts the median in $[\HK(G),(1+\eps)\HK(G)]$. Independence and a union bound give probability at most $2^k(1/8)^{k/2}=2^{-k/2}\le\upsilon$ that half the runs fail. The same argument applies to either TSP estimate.

Parallel copies share scans and use $O(k)$ times the space and work; completed copies remain idle. Sequential copies instead use $k$ times the passes and work, with $O(k)$ extra words for their outputs. A correct median need not come from a run with a feasible saved cover, so this amplifies only the value guarantee.

\subsection{Zero-cost edges}
One preliminary union--find pass groups vertices
connected by zero-cost paths. In later passes, replace each group by
its fixed representative, discard loops, and keep parallel copies.
This uses $O(n)$ words; return zero if one vertex remains.
For either the demand-one LP defining $z$ or the demand-two LP
defining $\HK(G)$, a cover of the contracted graph extends at the
same cost by assigning multiplicity two to every zero-cost edge and
zero to discarded positive-cost edges. Every cut splitting a group
crosses a zero-cost edge, so its total multiplicity is at least two;
all other cuts correspond to cuts of the contracted graph.
Conversely, restricting an original cover to edges between groups
gives a contracted cover of no greater cost. Thus both $z$ and
$\HK(G)$ are unchanged by the contraction.

\FloatBarrier
\section{Examples of stages and an outer round}\label{app:examples}
The first example explains the fixed search order for one tree, without
performing updates. The second follows a complete outer round, including
its thresholds, updates, and stopping test.

\subsection{Cut-search stages on six vertices}\label{app:stage-example}
Let $T$ consist of the path $v_0v_1v_2v_3v_4$ and the edge $f=v_0u$,
rooted at $v_0$. Write $e_i=v_{i-1}v_i$ for $1\le i\le4$.
The child subtrees below $e_1$ and $f$ have four and one vertices,
respectively, so $e_1$ is heavy. Each internal path vertex has one
child; hence the decomposition paths are $(e_1,e_2,e_3,e_4)$ and $(f)$.

Split the four-edge path into $(e_1,e_2)$ and $(e_3,e_4)$, then split
each half into single edges. Single-edge intervals have height zero,
the two halves have height one, and the full interval has height two.
The path $(f)$ has only height zero. Table~\ref{tab:stage-example}
lists the resulting stages. Each entry describes a cut by its tree
boundary, not by all its crossing edges in the input graph.

\begin{table}[!htbp]
\caption{Fixed cut-search stages for the six-vertex reference tree.
All five single-edge and ten two-edge descriptions occur exactly once.}
\label{tab:stage-example}
\centering
\begin{tabularx}{\textwidth}{@{}c>{\raggedright\arraybackslash}X>{\raggedright\arraybackslash}p{4.0cm}@{}}
\toprule
Stage & Cut descriptions & Reason\\
\midrule
0 & $\{e_1\},\{e_2\},\{e_3\},\{e_4\},\{f\}$ & One tree edge\\
1 & $\{e_1,e_2\},\{e_3,e_4\}$ & Same height-one interval\\
2 & $\{e_1,e_3\},\{e_1,e_4\},\{e_2,e_3\},\{e_2,e_4\}$ & Different halves of the height-two interval\\
3 & $\{e_i,f\}$ for $1\le i\le4$ & Different decomposition paths\\
\bottomrule
\end{tabularx}
\end{table}

For example, $e_2,e_3$ are adjacent, but $\{e_2,e_3\}$ belongs to
stage~2: the full four-edge interval is the smallest \emph{recursive}
interval containing them. Its middle shore is $\{v_2\}$, which contains
the separator $v_2$. By contrast, the stage-one pairs have middle shores
$\{v_1\}$ and $\{v_3\}$, neither containing that separator. This is the
distinction used in \cref{lem:root-prerequisites}.

Stage~3 includes $\{e_2,f\}$ even though its edges are not adjacent.
Its shore not containing the root is $\{v_2,v_3,v_4,u\}$, with tree
boundary exactly $\{e_2,f\}$. Thus a two-edge description need not
represent a connected shore.

\emph{Within an outer round.}
The algorithm processes stages~0--3 for this tree, updating only cuts
below the current search threshold, then processes the other reference
trees. The full families are listed here only for illustration; the
algorithm does not enumerate them. The next example follows all updates
in a complete round.
\FloatBarrier

\subsection{A complete outer round on five vertices}\label{app:example}
We fix valid outputs of the sparsifier and reference-tree procedures;
this example does not test their implementations. Decimals are for
readability; comparisons use the exact rational expressions.

\subsubsection{Graph and thresholds}
Take the unit-cost cycle on vertices $v_0,v_1,v_2,v_3,v_4$, and write $a_i:=v_{i-1}v_i$ for $1\le i\le4$ and $b:=v_0v_4$. Use the path $T=\{a_1,a_2,a_3,a_4\}$ rooted at $v_0$ as the reference
tree. Initially $w_e=q_e=1$ and every nontrivial cut crosses either two
or four cycle edges. The minimum cut is two. Every cut within $5/4$ of
this minimum crosses at most two edges of $T$, so $\{T\}$ is a valid
reference family. Each stored graph is the exact five-edge graph with
its current capacities, which satisfies every requested sparsifier bound.

Set $\eps=1/2$ and choose $\xi=1/128\le\eps/40$. There is one
decomposition path. Split $[a_1,a_4]$ into
$[a_1,a_2]$ and $[a_3,a_4]$. The stage families are $\mathcal F_0=\{\{a_1\},\{a_2\},\{a_3\},\{a_4\}\}$, $\mathcal F_1=\{\{a_1,a_2\},\{a_3,a_4\}\}$, $\mathcal F_2=\{\{a_1,a_3\},\allowbreak\{a_1,a_4\},\allowbreak\{a_2,a_3\},\allowbreak\{a_2,a_4\}\}$, and $\mathcal F_3=\varnothing$.
There is no cross-path pair because the tree has only one decomposition path.
For the full interval, the separator is $v_2$. Its middle shore
$\{v_p,\ldots,v_{q-1}\}$ contains $v_2$ exactly for the
four pairs in $\mathcal F_2$. The two pairs in $\mathcal F_1$ are
nonroot pairs and have already been processed when stage two starts.

Here $r=4$, $\zeta_0=1/2560$, and $\zeta=1/64000$. Initialization
divides the exact stored minimum by $1+\zeta_0$, giving $\lambda=5120/2561$, $L=5160/2561$, and $U=5280/2561$. Each stage multiplies the previous bound by $63999/64001$.
Thus $\lambda\simeq1.999219$, $L\simeq2.014838$, and
$U\simeq2.061695$. The initial checkpoint has $D_0=5$ and $Z_0=\frac{D_0}{\lambda}=\frac{2561}{1024}$. Table~\ref{tab:example-stages} gives the search thresholds and the groups
that will be processed. The cut descriptions determine their input-edge
boundaries exactly: $\{a_i\}$ crosses $a_i,b$, while
$\{a_i,a_j\}$ crosses only $a_i,a_j$.

\begin{table}[!htbp]
\caption{One outer round on the unit five-cycle. Thresholds are rounded for display; updates include each nonempty call's initialization. Every $\Lambda_j$ exceeds $L$.}
\label{tab:example-stages}
\centering
\begin{tabular}{@{}crrlr@{}}
\toprule
Stage $j$ & $\theta_j$ & $\Lambda_j$ & Retained group $B_j$ & Updates\\
\midrule
0 & $2.061662$ & $2.061630$ & all four singletons & 7\\
1 & $2.061598$ & $2.061566$ & $\{a_1,a_2\},\{a_3,a_4\}$ & 3\\
2 & $2.061534$ & $2.061501$ & none & 0\\
3 & $2.061469$ & $2.061437$ & none & 0\\
\bottomrule
\end{tabular}

\end{table}

\subsubsection{Updates and cut checks}
\emph{Stage zero: singleton cuts.}
Each singleton has initial capacity two, below $\theta_0$. The group
has four cuts with minimum crossing cost one, so $\alpha_S=\xi/4=1/512$.
A tree edge crosses one group cut, while $b$ crosses all four. The
initial row loads are therefore $h_{a_i}^0=\xi/4$ and $h_b^0=\xi$. The largest row load is already $\xi$. As long as all four cuts remain
active, the phase scalar is $\sigma_1=1$, and every later update has
the same loads. After $t$ updates, including initialization, $w_{a_i}=(1+\xi/4)^t$ and $w_b=(1+\xi)^t$. Every active cut has value $(1+\xi/4)^t+(1+\xi)^t$. The tests at the
last two iterations are $(1+\xi/4)^6+(1+\xi)^6\simeq2.059576<U$ and $(1+\xi/4)^7+(1+\xi)^7\simeq2.069738>U$.
Thus the call performs initialization and six noninitial updates, then
deletes all four cuts. Write $A_0:=(513/512)^7$ and $B_0:=(129/128)^7$. At the end, every tree edge has weight $A_0\simeq1.013752$, and
$b$ has weight $B_0\simeq1.055986$.

The stored record has four pairs $(\alpha_S,d(S))=(1/512,1)$ and
one phase pair $(\sigma_1,k_1)=(1,6)$, together with the tree pointer.
For example, replaying $b$ multiplies its previous weight by
$(1+\xi)(1+\xi)^6=(1+\xi)^7$. Initialization is a separate factor,
even though it happens to equal the later updates on this example.

\emph{Stage one and omitted cuts.}
Build a new stored graph with these updated weights. Both pairs in
$\mathcal F_1$ have value $2A_0\simeq2.027504<\theta_1$ and are
retained. The earlier single-edge cuts have value
$A_0+B_0\simeq2.069738>\theta_1$, as required by the path search.

This group has two cuts and $\alpha_S=\xi/2=1/256$. Each tree edge
crosses exactly one group cut, and $b$ crosses neither. Initialization
therefore adds load $\xi/2$ to each tree edge and zero to $b$.
The later phase scalar is $\sigma_1=2$, giving load $\xi$ to each
tree edge. After initialization and then $k$ noninitial updates, each
selected cut has value $2A_0(1+\xi/2)(1+\xi)^k$. For $k=1$ this is about $2.051326<U$; for $k=2$ it is about
$2.067352>U$. The call has three updates. Its record stores two pairs
$(\alpha_S,d(S))=(1/256,1)$ and $(\sigma_1,k_1)=(2,2)$.

Put $A_1:=A_0(257/256)(129/128)^2\simeq1.033676$. The new weights are $A_1$ on each tree edge and $B_0$ on $b$.
At stage two, all four pairs have value
$2A_1\simeq2.067352>\theta_2$, so the search returns no cuts.
Earlier pairs also exceed $\theta_2$. The final stage has no cross-path
pairs. Both empty searches belong to the round but need no \Focus call.

Table~\ref{tab:example-checkpoints} collects the states reached. It also
shows why a saved estimate need not correspond to the most recent weights.

\begin{table}[!htbp]
\caption{Weights, actual minimum cuts, and checkpoint objectives in the same example. The last row changes the lower bound, not the weights. Decimals are rounded; the verification script uses rational arithmetic.}
\label{tab:example-checkpoints}
\centering
\begin{tabular}{@{}lrrrr@{}}
\toprule
Checkpoint & Tree-edge weight & Weight of $b$ & $\mincut(q)$ & $D/\lambda_t$\\
\midrule
Initialization & $1$ & $1$ & $2$ & $2.500977$\\
After stage 0 & $1.013752$ & $1.055986$ & $2.027504$ & $2.556496$\\
After stages 1--3 & $1.033676$ & $1.055986$ & $2.067352$ & $2.596359$\\
After $\lambda\leftarrow L$ & $1.033676$ & $1.055986$ & $2.067352$ & $2.576232$\\
\bottomrule
\end{tabular}

\end{table}

\subsubsection{Ending the round and saving the estimate}
All capacities are at least $A_1$, and every cut crosses at least two
cycle edges. Two tree edges form a cut of value $2A_1$, so $\mincut(q)=2A_1>L$. The update $\lambda\leftarrow L$ is therefore valid. The cumulative
loads are $17/512$ on each tree edge and $7/128$ on $b$. Their maximum
is below $\Gamma=128\log5$, so the outer round ends normally, not because
of the stopping rule. The algorithm must continue to later outer rounds.
If the reference procedure supplied additional trees, their stages
would find no cuts below their search thresholds here, since already
$\mincut(q)>U$.

The objective $D/\lambda$ increases during the round. Raising $\lambda$
lowers it to about $2.576232$, still above $Z_0=2561/1024$, so the
initial checkpoint remains best. Its original bound is
$\lambda_0=5120/2561$, and its cover assigns
$y_e^*=2/\lambda_0=2561/2560$ to each of the five edges, giving
$\sum_e y_e^*=2561/512=5.001953125$. Both true optima are five, but
this accurate initial estimate does not justify early termination.
The checkpoint must retain $\lambda_0$: the new bound $L>2$ is invalid
for the initial capacities, whose minimum cut is two.

\FloatBarrier
\Needspace{10\baselineskip}
\section{TSP bounds and rounding}\label{app:tsp-transfer}

For an even vertex set $O$, an \emph{$O$-join}
is an edge set whose odd-degree vertices are exactly $O$.
Write $\tau_G(O)$ for its minimum cost.

The Edmonds--Johnson characterization~\cite{EdmondsJohnson,Schrijver} expresses $\tau_G(O)$ as the minimum of $\sum_e c_ey_e$ over nonnegative vectors satisfying $y(\delta(S))\ge1$ whenever $|S\cap O|$ is odd.
Every feasible cover in \eqref{eq:unit-cover} satisfies these constraints, so $\tau_G(O)\le z$ for every even $O\subseteq V$.

\begin{proof}[Proof of \cref{thm:MST-sandwich}]
Let $R$ be an MST and let $O_R$ be its set of odd-degree vertices,
which has even cardinality. Add a minimum $O_R$-join to $R$,
keeping both copies of any shared edge. The resulting multigraph
is connected and has even degrees, so an Euler tour visits every
vertex. By the join bound above, $\TSP(G)\le W(G)+\tau_G(O_R)\le W(G)+z$.
For the upper bound, add $W(G)\le\TSP(G)$ and $z=\HK(G)/2\le\TSP(G)/2$.
\end{proof}

\begin{proof}[Proof of \cref{thm:LP-transfer}]
The first term has the stated guarantee by
\cref{cor:MST-transfer}. For the second term, $\TSP(G)\le\gamma\HK(G)=2\gamma z\le2\gamma Z$. Also, $2\gamma Z\le\gamma(1+\eta)\HK(G)\le\gamma(1+\eta)\TSP(G)$.
Both terms are upper bounds, so their minimum gives the smaller
approximation factor.
\end{proof}

\emph{Why retain the MST term?}
Each tree edge is the sole edge of a cut, so on a tree $z=W$ and
$\TSP=2W$: the term $W+z$ is exact. On a unit-cost $n$-cycle,
vertex cuts give $z\ge n/2$, and $y_e=1/2$ is feasible, so $z=n/2$.
Since $W=n-1$ and $\TSP=n$, the ratio is $3/2-1/n$.
The MST term helps on individual instances but is asymptotically
$3/2$-tight.

\emph{A sharper general-metric factor.}
For general metrics, Gurvits--Klein--Leake's Lemma~13 and Corollary~14~\cite[p.~79:12]{GurvitsKleinLeake} give $\gamma_{\rm met}=3/2-2.18\cdot10^{-34}$, building on~\cite{KarlinKleinOG}.
Their calculation gives
$9.7(1.5\cdot10^{-9})^2\cdot10^{-17}
 =2.1825\cdot10^{-34}$.
After rescaling the estimation accuracy, the transfer gives
$\gamma_{\rm met}+\eps$. This uses their rounding bound and does
not improve it.

\emph{Rounding the reported cost.}
For rational costs, sum the retained MST-edge costs with upward rounding
to obtain $W\le W^+\le(1+\eta)W$. If $z\le Z\le(1+\eta)z$, then
$\TSP(G)\le W^++Z\le(1+\eta)(W+z)\le\tfrac32(1+\eta)\TSP(G)$;
the guarantee for $2\gamma Z$ is unchanged. Choose $\eta\le\eps/4$
for $0<\eps\le1/2$. Allowing a further factor $1+\eta$ for final
upward rounding, $\tfrac32(1+\eta)^2\le\tfrac32+\eps$ and
$\tfrac75(1+\eta)^2\le\tfrac75+\eps$, proving \cref{cor:tsp}
within the stated bounds. The lower-bound reduction needs no rounding:
its MST cost is a sum of polynomially bounded integers.

\emph{Comparison with max--min joins.}
Let $\mu:=\max_{O\subseteq V,\ |O|\text{ even}}\tau_G(O)$.
The parity and join bounds give $\TSP(G)\le W+\mu$ and
$\mu\le z\le\TSP(G)/2$. Iwata--Ravi's $2/3$-approximation~\cite{IwataRavi}
returns an even set $O$ with $2\mu/3\le a:=\tau_G(O)\le\mu$.
Thus $\TSP(G)\le W+\tfrac32a\le W+\tfrac32\mu\le\tfrac74\TSP(G)$.
This is a value comparison, not another streaming algorithm.
Estimating $z$ directly avoids this rounding loss.

\end{document}